\documentstyle[12pt,aasms4]{article}
\newcommand{\fluxunit}{$ \ {\rm erg}\ {\rm cm}^{-2}\ {\rm s}^{-1} \ $}

\slugcomment{Revised Oct. 12th}

\lefthead{Akiyama et al.}
\righthead{Optical Identification of the {\it ASCA} LSS}

\begin{document}

\title{Optical Identification of
the {\it ASCA} Large Sky Survey}

\author{Masayuki Akiyama\altaffilmark{1,2,10,11,12}, Kouji Ohta\altaffilmark{2,10,11},Toru Yamada\altaffilmark{3,11},
Nobunari Kashikawa\altaffilmark{4}, Masafumi Yagi\altaffilmark{4}, Wataru Kawasaki\altaffilmark{5,12},
Masaaki Sakano\altaffilmark{6,12}, Takeshi Tsuru\altaffilmark{6}, Yoshihiro Ueda\altaffilmark{7,10},
Tadayuki Takahashi\altaffilmark{7}, Ingo Lehmann\altaffilmark{8,10}, G\"unther Hasinger\altaffilmark{8},
and Wolfgang Voges\altaffilmark{9}}

\altaffiltext{1}{SUBARU Telescope, National Astronomical Observatory of Japan, 650 North A'ohoku Place, Hilo, HI, 96720, U.S.A}
\altaffiltext{2}{Department of Astronomy, Kyoto University, Kyoto 606-8502, Japan}
\altaffiltext{3}{Astronomical Institute, Tohoku University, Sendai 980-8578, Japan}
\altaffiltext{4}{National Astronomical Observatory of Japan, Mitaka, Tokyo 181-8588, Japan}
\altaffiltext{5}{Department of Astronomy, University of Tokyo, Tokyo 113-8658, Japan}
\altaffiltext{6}{Department of Physics, Kyoto University, Kyoto 606-8502, Japan}
\altaffiltext{7}{Institute of Space and Astronautical Science, Kanagawa 229-8510, Japan}
\altaffiltext{8}{Astrophysikalisches Institut Potsdam, An der Sternwarte 16, 14482 Potsdam, Germany}
\altaffiltext{9}{MPI f\"ur extraterrestrische Physik, Postfach 1603, 85740 Garching, Germany}
\altaffiltext{10}{Visiting Astronomer,
German-Spanish Astronomical Centre, Calar Alto, operated by the Max-Plank-Institute for
Astronomy, Heidelberg, jointly with the Spanish National Commission for Astronomy}
\altaffiltext{11}{Visiting Astronomer, University of Hawaii Observatory.}
\altaffiltext{12}{Research Fellow of the Japan Society for the Promotion
of Science.}

\begin{abstract}
\noindent
We present results of optical identifications
of the X-ray sources detected in the {\it ASCA} Large Sky Survey.
Optical spectroscopic observations were done for
34 X-ray sources which were detected with the SIS in the 2--7~keV band
above 3.5 $\sigma$. The flux limit corresponds to
$\sim 1\times10^{-13}$ \fluxunit in the 2--10~keV band.
The sources are identified with 30 AGNs,
2 clusters of galaxies, and 1 galactic star.
Only 1 source is still unidentified.

All of the X-ray sources that have a hard X-ray spectrum
with an apparent photon index of smaller than 1 in the 0.7--10~keV
band
are identified with narrow-line or weak-broad-line AGNs
at redshifts smaller than 0.5.
This fact supports the idea that absorbed X-ray spectra of narrow-line and
weak-broad-line AGNs make
the Cosmic X-ray Background (CXB) spectrum
harder in the hard X-ray band than that of a broad-line AGN,
which is the main contributor in the soft X-ray band.
Assuming their intrinsic spectra are 
same as a broad-line AGN (a power-law model with a photon index of 1.7),
their X-ray spectra are fitted with
hydrogen column densities of $\log N_{\rm H}({\rm cm}^{-2}) = 22 \sim 23$
at the object's redshift.
On the other hand,
X-ray spectra of the other AGNs are consistent with that of a
nearby type~1 Seyfert.
In the sample, four high-redshift luminous broad-line AGNs show a hard X-ray spectrum
with an apparent photon index of $1.3\pm0.3$.
The hardness
may be explained by the reflection component of a type~1 Seyfert.
The hard X-ray spectra may also be explained by absorption with
$\log N_{\rm H}({\rm cm}^{-2}) = 22 \sim 23$ at the object's redshift,
if we assume an intrinsic photon index of 1.7.
The origin of the hardness is not clear yet.

Based on the logN-logS relations of each population,
contributions to the CXB in the 2--10~keV band are estimated to be
9\% for less-absorbed AGNs ($\log N_{\rm H}({\rm cm}^{-2}) < 22$)
including the four high-redshift broad-line AGNs with a hard X-ray spectrum,
4\% for absorbed AGNs ($22 < \log N_{\rm H}({\rm cm}^{-2}) < 23$, without
the four hard broad-line AGNs), and
1\% for clusters of galaxies in the flux range from
$3\times10^{-11}$\fluxunit to $2\times10^{-13}$\fluxunit.
If the four hard broad-line AGNs are included in the absorbed AGNs,
the contribution of the absorbed AGNs to the CXB is estimated to be 6\%.

In optical spectra, there is no high-redshift
luminous cousin of a narrow-line
AGN in our sample. The redshift distribution of the absorbed AGNs
are limited below $z=0.5$ excluding the four hard broad-line AGNs,
in contrast to the existence of 15
less-absorbed AGNs above $z=0.5$.
The redshift distribution of the absorbed AGNs
suggests a deficiency of AGNs with column densities of
$\log N_{\rm H}({\rm cm}^{-2}) = 22$ to $23$ in the redshift range between 0.5
and 2, or in the X-ray luminosity range
larger than $10^{44}$ erg s$^{-1}$,
or both.
If the large column densities of the four hard broad-line AGNs are real, 
they could complement the deficiency of 
X-ray absorbed luminous high-redshift AGNs.

\end{abstract}


\section{Introduction}
\noindent
Since the discovery of the cosmic X-ray background (CXB) by
Giacconi et al. (1962) in the 2--6~keV band,
many efforts have been made to understand the origin of
the CXB.
Recently in {\it ROSAT\/} deep surveys,
70 -- 80\% of the CXB in the 0.5--2~keV band has been resolved
into discrete sources at a flux limit of $1 \times 10^{-15}$\fluxunit (\cite{has98}).
Within the flux level of the deep survey,
broad-line AGNs are the dominant population and
the main contributor to the CXB in the 0.5--2~keV band.
On the other hand, in the harder 2--10~keV band,
only $\sim 3$\%
of the CXB was resolved into discrete sources (\cite{pic82}) before {\it ASCA\/} surveys.
Broad-line AGNs have X-ray power-law spectra with a photon index of $\Gamma = 1.7$
in the 2--10~keV band (\cite{tur89}) which are
significantly softer than that of the CXB in that band
($\Gamma = 1.4 \sim 1.5$; \cite{gen95}; \cite{ish99}),
thus there must be
objects which have harder X-ray spectra than nearby broad-line AGNs
and contribute significantly to the CXB in the hard band.

Resolving the hard X-ray sky is a direct way
to reveal the nature of X-ray sources in the hard band.
{\it ASCA} GIS observations of three {\it ROSAT} Deep PSPC Fields down to the
$5\times10^{-14}$\fluxunit have been done so far (\cite{geo97}; \cite{boy98a}).
However, because of the large positional uncertainties of the GIS-selected X-ray sources,
a large fraction of hard X-ray selected sources is still unidentified;
the nature of hard X-ray sources and the difference from the {\it ROSAT} selected objects
are still unclear.

Based on the unified scheme of AGNs,
absorbed AGNs are proposed as candidates for the hard X-ray sources
(\cite{com95}; Madau, Ghisellini, \& Fabian 1994).
Because of absorption of soft X-ray photons by obscuring material,
they have harder X-ray spectra than type~1 AGNs.
To reproduce the CXB spectrum, it is argued that there are around three times more
absorbed AGNs
than non-absorbed AGNs in the universe and absorbed AGNs dominate the
CXB above 2~keV.
In consequence of the assumption, the existence of absorbed narrow-line QSO (so called type~2 QSO)
is expected.
From optical follow-up observations of X-ray surveys in the soft and hard bands, several luminous narrow-line AGNs
have been found at intermediate to high redshift universe
(e.g., \cite{sto82}; \cite{alm95}; \cite{oht96}; \cite{boy98b}; \cite{bar98}; \cite{sch98}).
On the other hand, existence of red {\it broad-line} QSOs whose red color suggests
absorption to its nucleus was reported from a radio survey (\cite{web95}).
Such a population was also identified in {\it ROSAT} surveys (\cite{kim99}) and {\it Beppo-SAX} surveys (\cite{fio99}),
but X-ray spectra, number densities and contributions to the CXB of both of these narrow-line and red broad-line QSOs
are not clear.

To reveal the nature of X-ray sources in the hard band,
studies on a well-defined sample are important.
We are now conducting an unbiased large and deep
survey with {\it ASCA} in the region near the north Galactic pole,
i.e., {\it ASCA} Large Sky Survey (hereafter LSS;
\cite{ino96}; \cite{ued96}; \cite{ued98}; Ueda et al. 1999a (Paper I)).
The flux limit of the LSS ($\sim$ 1 $\times 10^{-13}$ \fluxunit\
in the 2--10~keV band) is 100 times deeper than
the {\it HEAO1\/}~A2 survey, which was the deepest
systematic survey in the hard band (\cite{pic82}) before {\it ASCA}.
We have surveyed 7.0 deg$^2$ and 5.4 deg$^2$ with the GIS and the SIS detectors,
respectively.
Combining the data from the GIS and the SIS,
we detected 44 sources in the 2--10~keV band
with the following criteria:
1) the significance of summed count rate of the 
GIS and the SIS should exceed 4.5, and
2) the significance of either the GIS or the SIS 
should also exceed 3.5 (\cite{ued99}).
They correspond to 20 -- 30\% of the CXB in this band.
The advantages of the sample are 1) the survey area as a function of limiting count rates
was determined well by simulations, 2) accurate X-ray positions were determined by
using the SIS whose resolution is higher than the GIS
and by correcting
the temperature-dependent misalignment between the focal plane detectors and the attitude sensors,
and 3) the X-ray spectrum of each source was determined
by fitting power-law model with the Galactic absorption  
to the GIS and the SIS data simultaneously in the 0.7--10~keV band (\cite{ued99}).
Determining their X-ray spectra is important
not only to know their X-ray properties
but also to know the flux limit of the survey, because the limit varies
with the X-ray spectrum of each source.
The average of the apparent photon index of the 36 X-ray sources detected
in the flux range between 0.8 $\times 10^{-13}$
and 4 $\times 10^{-13}$ \fluxunit\
is $\Gamma = 1.49 \pm 0.1$ (\cite{ued99}),
which is significantly harder than the spectra of
X-ray sources detected in shallower surveys in the
2--10~keV band and close to that of the CXB.
Identification of these
sources is clearly important to understand the nature of hard X-ray sources
and the origin of the CXB.

In this paper, we report results of optical identifications of
X-ray sources detected in the hard band with the SIS in the {\it ASCA} LSS.
The sample definition and
selections of optical candidates are discussed in Section 2,
results of optical spectroscopy for the selected candidates
and reliability of the identifications are presented in Section 3, and
optical and X-ray spectroscopic properties of identified objects are described in Section 4.
The contribution of the population to the CXB and
the redshift distribution of the AGNs are discussed in Section 5 and 6, respectively.
In Section 7, we present multi-wavelength properties of the identified AGNs.
Throughout this paper, we use $q_0=0.5$ and $H_0=50$ km s$^{-1}$ Mpc$^{-1}$.
We call each X-ray source with the exact name and the identification number,
like AX~J132032+3326(227), for convenience.

\section{Observations}

\subsection{The X-ray Survey Observations and the Sample Definition}
\noindent
To minimize the effect of the galactic absorption and
contamination from bright X-ray sources,
the survey area was defined as a continuous region
of $\sim$ 5 degree$^{2}$ near the north Galactic pole, centered at
$\alpha$=13$^{\rm h}$14$^{\rm m}$,
$\delta$=31$^{\circ}$30$'$ (J2000).
Seventy-six pointings of the survey-observations were done from
Dec. 1993 to Jul. 1995.
These exposures were designed to evenly cover the whole survey region
with a 20 ksec effective exposure of the SIS.
The source detections were done with the SIS data
in the 0.7--7~keV, 0.7--2~keV, and 2--7~keV bands and the GIS data
in the 0.7--7~keV, 0.7--2~keV, and 2--10~keV bands.
For details of the survey observation, source extraction, and spectral fitting,
including the survey region and the position of the sources on the sky, see \cite{ued99}.

In this paper, we concentrate on the 34 X-ray sources detected
with the SIS in the 2--7~keV band above 3.5$\sigma$
(hereafter, the SIS 2--7~keV 3.5$\sigma$ sample).
Table \ref{area} shows
the survey area as a function of limiting count rates for the sample.
The typical and the deepest limiting count rates of the sample are
2 counts ksec$^{-1}$ and 1.2 counts ksec$^{-1}$, respectively.
They correspond to $1.8\times10^{-13}$\fluxunit and $1.1\times10^{-13}$\fluxunit
in the 2--10~keV band for an X-ray source with a power-law spectrum with
a photon index of 1.7.
All X-ray sources in the SIS 2--7~keV 3.5$\sigma$ sample 
have the significance level larger than 4.5 in summed count rate of
the GIS and the SIS in both of the 0.7--7~keV and 2--7~keV bands
and the positional uncertainties of such sources 
were estimated to be $0.\!^{\prime}6$ 
in radius with the 90\% confidence level (\cite{ued99}).
The number of spurious sources was estimated to be at most
a few percent (\cite{ued99}),
thus less than 1 spurious source was expected to be in the SIS 2--7~keV 3.5$\sigma$
sample.

\placetable{area}

\subsection{Optical Imaging Observations and
Selections of Optical Counterpart Candidates}
\noindent
AGNs are the most plausible optical counterparts for the majority of the
X-ray sources, thus at least we have to reach the optical
flux limit which is converted from the deepest X-ray flux limit based
on X-ray-to-optical flux ratio of AGN.
If we assume power-law spectra with
an X-ray photon index of 1.7 and an optical energy index of $-0.5$
together with the X-ray-to-optical flux ratio of
AGNs identified in the {\it ROSAT\/} Deep Survey
(\cite{sch98}), the expected optical magnitude for the optical counterpart of an X-ray source with
$1.1\times10^{-13}$\fluxunit in the 2--10~keV band is $R=16\sim21$ mag.

Candidates of optical counterparts were mainly selected from the APM catalog
which is obtained from scans of glass copies of
the Palomar Observatory Sky Survey plate (\cite{mcm92}).
However, the limiting magnitude of the catalog is $R\cong20$ mag and
is not deep enough to pick up optical counterparts for faint X-ray sources.
To complement the depth of the data,
we made imaging observations of the LSS region at the KISO 1.05m
Schmidt telescope in March and April 1994.
In these observations, we used the mosaic CCD camera (\cite{sek92})
which is made up of 15 1024$\times$1024 CCD chips and
covers 2$\times$5 degree$^2$ with 15 shots.
The spatial resolution in the setup was $0.\!^{\prime\prime}75$ pix$^{-1}$.
Images were taken in the $R$ band with an exposure time of 20 minutes.
The reduction of the data was done by
the usual method for optical imaging data.
The weather condition during the observation was
neither stable nor photometric and
the limiting magnitude changed from field to field. The typical
seeing was $4^{\prime\prime}$ (FWHM) and
the typical limiting magnitude was $R\cong21$ mag, about
one magnitude deeper than the APM data.

There are several optical objects within each error circle of X-ray source above $R=21$ mag.
Two X-ray sources (AX~J132032+3326(227) and AX~J130748+2925(002))
show clear excesses of galaxies in and around their error circles (see
Figure \ref{plate}).
These two objects have been cataloged as candidate clusters of galaxies,
Abell 1714 (Abell, Corwin, \& Olowin 1989) and Zwcl 1305.4+2941 (Zwicky, Herzog, \& Wild 1961), respectively.
Zwcl 1305.4+2941 was also detected in the {\it Einstein} Medium Sensitivity
Survey and its redshift was determined to be z=0.241 (\cite{sto91}).
We took deeper and better resolution images
with a Tektronix 2048$\times$ 2048 CCD on the University of Hawaii $88^{\prime\prime}$ telescope
on 1995 March and 1996 April in the $R$ and $I$ band and confirmed the excess of galaxies. Thus, we identified the two sources with clusters of galaxies.
For other sources,
active galaxies are the most plausible counterparts.
From optical objects within $0.\!^{\prime}8$, 
which is slightly larger than the estimated 90\% confidence error
radius ($0.\!^{\prime}6$), of each X-ray source,
we selected targets for spectroscopy, using 1) deeper X-ray follow-up data in
the soft and hard band, 2) radio emission, and 3) blueness of UV or optical color which
indicate existence of an activity in a object.

\subsubsection{Deep Follow-up Observations in X-ray band}
\noindent
To pinpoint optical counterparts of X-ray sources,
20 ksec follow-up observations were made with the {\it ROSAT} HRI in 2 fields
in December 1997.
We selected two fields centered at $\alpha$=13$^{\rm h}$14$^{\rm m}36^{\rm s}$,
$\delta$=32$^{\circ}$01$'$12$''$ and $\alpha$=13$^{\rm h}$12$^{\rm m}29^{\rm s}$,
$\delta$=31$^{\circ}$13$'$12$''$ (J2000) with a radius of $19'$,
to cover as many {\it ASCA} sources as possible.
We summarize the data reduction and the source detection in Appendix A.
All of the X-ray sources in the SIS 2--7~keV 3.5$\sigma$ sample
(AX~J131521+3159(136), AX~J131407+3158(127), and AX~J131327+3155(121) in the former field,
and
AX~J131249+3112(096), AX~J131128+3105(080), and AX~J131321+3119(103) in the latter field)
except one (AX~J131345+3118(104)) in these fields were detected by the HRI and
their optical counterparts were pinpointed thanks to
the $5^{\prime\prime}$ positional accuracy of the HRI.
On the position of the missed source (AX~J131345+3118(104)), there is an X-ray peak in the
HRI image, though the significance level is slightly lower than the detection limit.
Since there is also an optical object at the position,
the X-ray source is pinpointed.
It should be noted that
there is no hard X-ray source which has an X-ray spectrum with
an apparent photon index smaller than 1.0 in these HRI observed fields.

Deep 40 $\sim$ 100 ksec pointing observations for 4 sources, which have hard X-ray spectra,
(AX~J131501+3141(119), AX~J131551+3237(171), AX~J131210+3048(072),
and AX~J130926+2952(016)) were made by {\it ASCA} and
more precise positions and X-ray spectra were obtained
(\cite{sak98}; \cite{sak99}; \cite{ued99b}).
The error circles for these sources were estimated to be less than
$0.\!^{\prime}6$ at a 90\% confidence level.

By the {\it ROSAT} PSPC, 15 and 13 {\it ASCA} LSS
sources were detected in the {\it ROSAT} PSPC 
All-Sky Survey (\cite{vog99}) and 
pointing observations (Voges, private communication), respectively.
9 of them were detected in both of the observations,
thus the X-ray positions of 19 {\it ASCA} sources were determined accurately.
In details of the cross-identification, see Appendix B.

\subsubsection{Cross-correlation with FIRST radio source catalog}
\noindent
In the course of the optical counterpart selection,
we examined the distribution of FIRST radio sources
around LSS X-ray sources to evaluate the cross-correlation
between radio and X-ray sources.
The FIRST survey is a radio source survey conducted with the Very Large Array
in the 1.4 GHz band with a $5\sigma$
limiting flux of 1 mJy (\cite{bec95}).
There are 17 radio sources within $0.\!^{\prime}5$ from the centers of the
X-ray sources; by contrast no radio source exists between $0.\!^{\prime}5$
and $1^{\prime}$ from them.
Based on the surface number density of detected radio sources in the
LSS field, the contamination of a radio source which is not an
counterpart of an X-ray source
is expected to be less than 1 source
for the whole 34 X-ray sources within $0.\!^{\prime}5$.
Thus, the 17 radio sources are likely radio-counterparts of X-ray sources.
In details of the cross-identification, see Appendix B.
Three X-ray sources have two to four radio sources within $0.\!^{\prime}5$.
They are thought to be radio-loud objects with radio lobes or clusters
of galaxies.
One of the three X-ray sources is already identified with
a cluster of galaxies (AX~J132032+3326(227)).
In summary,
35\% (12/34) of the SIS 3.5$\sigma$ sample are detected
in the FIRST survey and an optical counterpart is pinpointed
thanks to the good positional accuracy ($1^{\prime\prime}$) of the FIRST catalog, except for AX~J131639+3149(137).
For the X-ray source,
there is no optical object within an error circle of the FIRST radio source
(see notes in Section 3.1).
In the error circle of AX~J131529+3117(110), there is a radio
peak with a flux of 0.63 mJy which is slightly lower than
the detection limit of the FIRST. Because there is an optical object
at the position, we identified the optical object with the 
counterpart of the radio peak and the X-ray source.

The high fraction of radio-detected X-ray sources in comparison with
other results (e.g., 10\% in \cite{DeRu97}) is due to
a match between the flux limit of the {\it VLA} FIRST 1.4 GHz survey and
that of the LSS;
the ratio of the flux limits is one order of magnitude larger than
the typical radio-to-X-ray flux
ratio of {\it radio-quiet} broad-line AGN (\cite{elv94}) but within the
scatter of the ratio of the {\it radio-quiet} population.
Thus, not only radio-loud AGNs but also some radio-quiet AGNs are
included in the LSS-FIRST sample (see Section 7.2).
It is worth noting that radio emission is transparent against
obscuring material, thus obscured AGNs are detected
in the radio wavelength as well as in the hard X-ray band, in an unbiased way.

\subsubsection{UV and Optical Color}
\noindent
Based on the results of the optical identification in
soft X-ray surveys,
a broad-line AGN is a plausible optical counterpart for a part of X-ray sources
detected in the hard band
and they can be selected by a blueness of their UV or optical color
if they are at redshift smaller than 3.
There are two UV-excess surveys which cover the southern half
($\delta < 32^{\circ}10'$) of the LSS.
These are Usher (1981) and Moreau \& Reboul (1995).
Objects with an $U-V$ color bluer than 0.0 down to $B=20$ mag and 0.1 down to $V=20$ mag
are cataloged in the former and the latter, respectively.
8 out of 20 X-ray sources in the area have one or two UV-excess object(s)
within $0.\!^{\prime}8$ in total. 
6 are also detected in the X-ray
follow-ups with {\it ROSAT} HRI or in the FIRST survey.
Considering the number density of UV-excess objects in Moreau \& Reboul (1995),
the expected number
of contamination within $0.\!^{\prime}8$ from the center of the 20 X-ray
sources is calculated to be 
0.5, thus all 8 correlated objects can be considered as
plausible optical counterparts of X-ray sources.
Such a low probability of contamination indicates that there is no serious
bias to selecting an UV excess object which is not a true 
optical counterpart of an X-ray source.
Additionally, two galaxies in the $0.\!^{\prime}8$ radii 
of AX~J131805+3349(233) are cataloged as UV-excess
galaxies in the KUG catalog (\cite{tak86}).

To select candidates of broad-line AGNs with fainter 
magnitude and to cover the
whole LSS area,
we also picked up objects with $B-R < 1.0$ mag
as a broad-line AGN
candidates in the whole area,
assuming a typical color of 
broad-line QSOs ($B-R=0.52$ mag; \cite{ric80}) with a
dispersion of distribution ($\Delta B-R=0.2$ mag) and an error of photometry
($\Delta B-R=0.3$ mag).
The APM $O-E$ color was converted to the $B-R$ color with
$(O-E)_{\rm APM} = 1.135 \times (B-R)$ (\cite{eva89}).
19 X-ray sources have one or two 
optical object(s) which meet the above criteria
within $0.\!^{\prime}8$.
5 of them are also detected in either the X-ray
follow-ups with {\it ROSAT} HRI or in the FIRST survey.
Within $0.\!^{\prime}6$ and
between $0.\!^{\prime}6$ and $0.\!^{\prime}8$
from the 34 X-ray sources, there are 19 and 6 blue objects, respectively.
Based on the number density of objects with $B-R$ color
bluer than 1.0 in the APM catalog, the expected number of contamination of 
blue objects which are not X-ray source is estimated to be
6 and 5 in the inner circles and outer annuli, respectively.
Thus, two thirds of the blue objects in the inner circles are
expected to be true optical counterpart of X-ray sources, on the other hand,
most of the blue objects in the outer annuli are
thought to be contamination of non-X-ray objects. 

Almost all (5 out of 6) the hard X-ray
sources which have a photon index smaller than 1.0
have no such blue object within their error circles;
therefore an optical counterpart
different from a broad-line AGN is suggested.
This tendency is the same as the previous results reported in Akiyama et al. (1998b),
but the new result is more significant than the previous one,
because we concentrate on brighter X-ray sources whose
positions and X-ray photon indices were determined well and
more accurate X-ray positions were obtained for each X-ray source
by correcting for the temperature effect in the satellite acquisition
as mentioned in Section 2.1.

In summary, except for 2 clusters of galaxies,
24 sources were pinpointed by X-ray follow-ups with
the {\it ROSAT} HRI or PSPC or
radio detections, 5 more sources have only one object which is
selected by its blue optical color,
and the remaining 3 sources have no optical counterpart which
meets the above criteria.
For these 3 sources, we picked up the brightest
optical object within each error circle as a primary target.
The list of the selected objects
is shown in Table \ref{id_table} and,
in finding charts of Figure \ref{plate},
they are labeled with ``A'', ``B'', and ``C'' in order of distance from the X-ray centroid.
Additionally, we listed some objects
which were not selected by the above methods
but observed in the optical spectroscopy.
These objects were indicated with a label ``Z''.

\placetable{id_table}

\subsection{Optical Spectroscopy}
\noindent
We made spectroscopic observations for
the optical-counterpart candidates
which are selected based on the above criteria with the highest priority.
Spectroscopic observations were made with
the University of Hawaii $88^{\prime\prime}$ telescope
on March 1998,
except for the hardest source
(AX~J131501+3141(119)),
a candidate of galactic star (AX~J131850+3326(219)), and
the two clusters of galaxies.
We used the Wide Field Grism Spectrograph with a
grating of 420 rulings mm$^{-1}$ and the
blaze wavelength of 6400\AA.
The spatial resolution was $0.^{\prime\prime}354$ pixel$^{-1}$ and the
seeing during the observation was $0.^{\prime\prime}8 \sim 1^{\prime\prime}$.
A slit width of $1.^{\prime\prime}2$ was used.
The spectral coverage ranges from 4000{\AA} to 9000{\AA} and
the spectral resolution, which is measured by the
HgAr lines in comparison frames and night-sky lines
in object frames, is 12{\AA} (FWHM).
We also took imaging data of each X-ray source with a better angular
resolution without filter for finding charts.

We have also done spectroscopic observations
with the 3.5m telescope at Calar Alto observatory
for a candidate galactic star (AX~J131850+3326(219)) and
three objects (AX~J131831+3341(228), AX~J130840+2955(014), and
AX~J131639+3149(137)) which were already observed in the previous run.
We used the MOSCA instrument in a single-slit mode with a g250 grating
which has 250 rulings mm$^{-1}$ and a blaze wavelength of 5700{\AA}.
The spectral coverage ranges from 4000{\AA} to 8000{\AA}.
In the configuration, the sampling was 5.95 {\AA} pixel$^{-1}$.
A slit width of $1.^{\prime\prime}5$,
which was the same as the FWHM of the seeing,
was used. Thus, the spectral resolution was 22{\AA} (FWHM),
which was estimated by measuring widths of night sky lines in object frames.
The spatial resolution was $0.^{\prime\prime}32$ pixel$^{-1}$.

For the hardest X-ray source (AX~J131501+3141(119)),
we made spectroscopic observations
with the Kitt Peak National Observatories Mayall 4m and 2.1m telescopes.
For details of the observations, see Akiyama et al. (1998a).

The data were analyzed using IRAF.\footnote{
IRAF is distributed by the National Optical Astronomy Observatories,
which is operated by the Association of Universities for
Research in Astronomy, Inc. (AURA) under cooperative 
agreement with the National Science Foundation.}
After bias subtraction, flat-fielding, and wavelength calibration,
optimum extraction method by {\bf apextract} package was used to
extract one dimensional spectral data from the two dimensional
original data.
For the UH data,
flux calibrations were done with Feige 34.
The flux calibrations did not work well
in the wavelength larger than 7000{\AA} and for
some data in the wavelength shorter than 5000{\AA} taken at very large zenith distances.
For the Calar Alto data, flux calibrations were done with HD84937.
The data were affected by fringes in the wavelength longer than 7000\AA.
Spectral fitting for emission lines was made by the $\chi^2$ minimization method
with {\bf specfit} command in {\bf spfitpkg} package
in the IRAF.
FWHMs of line widths were deconvolved by the spectral resolution shown above.

\section{Results and Reliability}

\subsection{Results of Optical Identification}
\noindent
Many (25 out of 34) X-ray sources have one optical-counterpart candidate which
shows broad permitted-emission lines whose FWHM is larger
than those of forbidden-emission lines or 1000 km s$^{-1}$ in their optical spectra. The detected broad lines and line widths are summarized in 
Table \ref{spe_table}.
This suggests that most of the X-ray sources originate from AGNs.

In the remaining sources,
5 X-ray sources have only an object with narrow-emission lines.
For 4 of them (AX~J131758+3257(195), AX~J131551+3237(171),
AX~J131501+3141(119), AX~J131210+3048(072)),
both the H$\alpha$ and H$\beta$ regions were observed.
We classified these objects, using
the [NII] 6583\AA-to-H$\alpha$ and [OIII] 5007\AA-to-H$\beta$ line ratios
(\cite{ost89}).
All of them show strong [NII] 6583{\AA} line as well as an H$\alpha$ line
and they fall in the region occupied by AGNs.
For the remaining one object, AX~J130840+2955(014),
spectroscopic observations cover only the H$\beta$ emission line region,
because of atmospheric absorption and fringes.
It has a large [OIII] 5007\AA-to-H$\beta$ flux ratio and
the existence of an AGN is suggested, but the ratio is
not large enough to exclude possibility of star-forming galaxy.
However, the spectrum shows optical continuum which is dominated by old stellar
populations and there is no indication of star-formation activity.
Thus, the object is also identified with AGN.

We list notes on problematic cases, below.

\begin{description}
\item[AX~J131054+3004(037)]
The FIRST source (A in Figure \ref{plate}) in the error circle was identified with a
narrow-line AGN at a redshift of 0.245. There is also a blue object within the error circle (B)
and it was identified with a broad-line AGN at a redshift of 1.577.
Because most other X-ray sources whose X-ray spectra are similar
to this X-ray source were identified with broad-line AGNs,
we assigned the broad-line AGN as the optical counterpart of the X-ray source.
It is possible that the narrow-line AGN or both is the origin of the X-ray source.

\item[AX~J131639+3149(137)]
At the position of the FIRST source in the error circle
($\alpha$=13$^{\rm h}$16$^{\rm m}38.9^{\rm s}$,
$\delta$=31$^{\circ}$49$'$57.8$''$),
no optical object brighter than $R=22.5$ mag can be seen.
The nearest object, which is separated by 14$''$ from the radio position and
has a blue color,
was identified with a broad-line AGN at z=0.622.
The projected distance between radio and optical
source corresponds to 108 kpc at the redshift.
If we assume that the radio emission is originated from the object,
the radio-to-X-ray flux ratio corresponds to that of radio-loud objects (see Section 7.2).

\item[AX~J131832+3259(199)]
The brightest stellar object (A in Figure \ref{plate}) in the
$0.\!^{\prime}8$ error circle
was identified with a G-type star.
Based on an X-ray-to-optical flux ratio of a G-type star, this object cannot
be the origin of the X-ray source.
This source is not identified yet.

\end{description}

As a result,
34 X-ray sources in the SIS 2--7~keV 3.5$\sigma$ sample were identified
with 30 AGNs including 5 objects with only narrow-emission lines,
2 clusters of galaxies,
and 1 galactic star. Only 1 source is still unidentified.
There are 2 X-ray sources which have HII-region like galaxies in their
error circles. But they also have an optical counterpart candidate
with AGN activity, thus we identified the X-ray sources with the AGNs.
The resulting identification and classification for each X-ray source
is indicated with bold characters in Table \ref{id_table}.
The optical spectra of the identified objects are shown in Figure \ref{plate}.
Detected broad-lines in each object are listed in Table \ref{spe_table}.
Physical parameters of the identified objects are summarized
in Table \ref{flux_table}.
In the next 2 subsections, we discuss the reliability of the
identifications.

\placefigure{plate}
\placetable{spe_table}
\placetable{flux_table}

\subsection{Contamination}
\noindent
Figure \ref{coord} shows the distribution of optically identified objects 
in the X-ray error circles.
Almost all X-ray sources were identified with optical objects within $0.\!^{\prime}6$,
though we picked up the candidates within $0.\!^{\prime}8$.
This result confirms the estimated 90\% confidence error radius of
$0.\!^{\prime}6$.

We have checked the reliability of our identification by estimating
chance contaminations in the error circle.
Based on number counts of optically-selected broad-line AGN brighter than
$B=21.0$ mag with a redshift smaller
than 3 (\cite{har90}), the expected number of
chance contaminations of broad-line AGN within 34 error circles is 0.29 and
is negligible.
For narrow-line AGNs,
3/5 of them, are pinpointed by the FIRST radio survey and
chance contamination is very small for the sample as mentioned above.
Therefore contamination of an object which is not an
X-ray source is very low.

\placefigure{coord}

\subsection{X-ray-to-optical Flux Ratio of the Identified Objects}
\noindent
An X-ray-to-optical flux ratio is one of useful tools to check the
reliability of identification.
To compare the X-ray-to-optical flux ratios of identified objects with
$\log f_{\rm X}/f_V$ of the {\it Einstein} Medium Sensitivity Survey (EMSS) sample
(\cite{sto91}),
we plot the relation between optical magnitude and hard X-ray flux
in Figure \ref{opt_x1}.
To plot the equal $\log f_{\rm X}/f_V$ lines,
we converted the $V$ band magnitude to the $R$ band magnitude and
0.3--3.5~keV flux to 2--10~keV flux for the EMSS sample,
using a $V-R$ color of $0.22$ mag, which is equivalent to a
typical optical energy index of
broad line AGNs ($-0.5$) and a photon index of 1.7 in the 0.3--10~keV band,
respectively.
The $\log f_{\rm X}/f_V$ values of
the EMSS AGNs, clusters of galaxies, and galactic G-type stars
are distributed $-1.0$ to $+1.2$, $-0.5$ to $+1.5$, and $-4.3$ to $-2.4$
(\cite{sto91}), respectively.
In the fainter flux range, the AGNs detected in the {\it ROSAT}
HRI Lockman Hole survey also occupy the same value of
$\log f_{\rm X}/f_V$ (\cite{sch98}).
The $\log f_{\rm X}/f_V$ values of the identified AGNs, clusters of
galaxies, and a galactic star in our sample are consistent with the distribution
of the EMSS sample.
The consistency of $\log f_{\rm X}/f_V$ with the EMSS objects supports
the reliability of
the identification in the LSS sample.

However,
there are two AGNs
which have values of $\log f_{\rm X}/f_V$
slightly outside of the EMSS sample.
One AGN (AX~J131831+3341(228)) has a $\log f_{\rm X}/f_V$ value of $+1.3$,
which is X-ray louder than the EMSS AGNs.
The AGN has a strong narrow [OIII] 5007{\AA} line compared with the 
broad H$\beta$ line.
Thus,
the large $\log f_{\rm X}/f_V$ value
might be explained by an optical absorption of its nucleus.
The other
AGN (AX~J131805+3349(233)),
whose $\log f_{\rm X}/f_V$ is smaller than $-1.0$, is dominated by its host
galaxy in the optical light (see Figure \ref{plate})
and this component makes the object
optically brighter in comparison with other AGNs.
The lower limit of $\log f_{\rm X}/f_V$ on the optical counterpart of the
unidentified source, AX~J131832+3259(199), is $\sim +1$.
This value is consistent with that of the EMSS AGNs.

\placefigure{opt_x1}

\section{Optical and X-ray Spectral Properties of the Identified AGNs}

\subsection{Strength of Broad Emission Lines}
\noindent
To examine the strengths of broad-lines in comparison with narrow-lines,
we measured the equivalent width ratios
of {\it broad plus narrow permitted lines}-to-{\it a narrow forbidden line}.
For objects at low redshifts,
we measured the equivalent width ratios of
H$\alpha$-to-[NII] 6583{\AA} and H$\beta$-to-[OIII] 5007\AA.
For objects at redshifts between 0.3 and 0.75, we could measure only
H$\beta$-to-[OIII] 5007{\AA} ratio
(we could not measure the ratio of AX~J131639+3149(137),
because of the fringe effect, thus the object is included in the
higher redshift sample).
The results are summarized in Table \ref{spe_table} and shown in Figure \ref{RAB}.
The objects with strong broad H$\alpha$ and H$\beta$ lines
occupy the upper-right region of the figure.
Most objects at redshifts lower than 0.3 do not show
significant broad component in the H$\beta$ line
and are placed in the lower part.
About half of them show a broad component in the H$\alpha$ line and
are distributed in the lower-right region.
In the right box,
among objects with redshifts between 0.3 and 0.75,
most of the objects show significant broad component
in the H$\beta$ line.
In this redshift range,
the fraction of objects with large H$\beta$-to-[OIII] 5007{\AA} ratio is
larger than the lower redshift sample in the left box.
Based on the criteria used in Winkler (1992),
the upper and lower
boundary of the H$\beta$-to-[OIII] 5007{\AA} ratio 
distribution in
our sample correspond to type~1.2 and type~1.8 $\sim$ 2
Seyfert, respectively.

All 11 objects at redshift above 0.75 and AX~J131639+3149(137)
show broad MgII 2800{\AA} or MgII 2800{\AA} and CIII] 1909{\AA} lines.
The equivalent widths and the line widths of their MgII 2800{\AA} lines
are summarized in Table \ref{spe_table}.
Their equivalent widths are consistent with
those of a composite spectrum of optically-selected QSOs 
($50\pm29${\AA}; \cite{fra91}).
AX~J131707+3237(175) and AX~J131021+3019(039) have a narrow-emission component
in the MgII 2800{\AA} and CIII] 1909\AA, respectively. Their line widths are 
as large as 1400 km s$^{-1}$ in FWHM
and significantly larger than line widths of typical narrow-lines.
Thus, there is no high-redshift luminous
cousin of a narrow-line AGN in our sample.
Such a deficiency of luminous narrow-line AGN
is also reported in a radio selected
sample of AGN (e.g., \cite{law91}) and a far-infrared selected sample (e.g., \cite{bar95}).

\placefigure{RAB}

\subsection{X-ray Spectral Properties and the Effect of Absorption}
\noindent
The distribution of the apparent photon indices of the identified
objects determined in the
0.7--10~keV band as a function
of redshift is shown in Figure \ref{z_pi}.
All X-ray sources which have an apparent photon index smaller than 1
are identified with AGNs at redshifts smaller than 0.5.
In the previous Figure \ref{RAB}, we mark the X-ray sources which have
photon indices smaller than 1 with dots.
AX~J131551+3237(171), AX~J131501+3141(119), AX~J131210+3048(072), and
AX~J130840+2955(014) were identified with AGNs which show no 
significant broad line
in H$\alpha$ and H$\beta$.
Among them, the hardest source, AX~J131501+3141(119), is identified with a type~2
Seyfert at z=0.072 (\cite{aki98a}; \cite{sak98}).
The remaining hard X-ray source (AX~J130926+2952(016)) was identified with an
AGN which shows a weak broad H$\beta$ line and has a small H$\beta$-to-[OIII]5007{\AA}
ratio. Thus, the object corresponds to a type~1.8 Seyfert (\cite{win92}).
These identifications support the picture (e.g., \cite{com95})
that the absorbed X-ray spectrum of narrow-line
and weak-broad-line AGNs make the CXB spectrum harder than
X-ray spectrum of broad-line AGN in the hard band.
There still remains one hard source with a photon index of 0.58
unidentified.

\placefigure{z_pi}

Photon indices of other AGNs are distributed in the range from 1.2 to 2.1.
There are three broad-line AGNs with an apparent photon index
as small as 1.4 at intermediate to high redshifts.
A dashed line in the figure shows the expected change of
the {\it apparent} photon index in the 0.7--10~keV band
with redshift for a type~1 Seyfert.
It is derived by fitting a power-law model to the
redshifted average X-ray spectrum of type~1 Seyferts
(\cite{gon96})
in the 0.7--10~keV band.
Because of the reflection component, the
observed photon index of type~1 Seyfert in the 0.7--10~keV
band is expected to get harder to $z \sim 2$ and
to reach a minimum photon index of 1.4 at a redshift of 2.5.
The photon index distribution of the broad-line AGN sample is
consistent with this line.
Thus, the hardness of the apparent photon indices of the
three high-redshift broad-line AGNs may be explained by the 
existence of the reflection component.
It should, however, be noted that the existence and the strength
of a reflection component
in AGNs with a luminosity of 10$^{45}$ erg s$^{-1}$
is not confirmed, so far.

For the AGNs, we also fitted their X-ray spectra with
an absorbed power-law model, assuming absorbing matter at the
object's redshift and an intrinsic photon index of 1.7.
The results are shown in column 8 of Table \ref{flux_table}
and shown in Figure \ref{lum_nh}
as a function of X-ray luminosities.
The upper boundary of the distribution of the
column density is consistent with an estimated limit on the column
density in the LSS sample (see, Section 6).
Hard X-ray sources which were identified with narrow-line or weak-broad-line AGNs
at redshift smaller than 0.5
are fitted with column densities of $\log N_{\rm H}({\rm cm}^{-2}) = 22 \sim 23$.
Most of the other AGNs are fitted with column densities
less than $\log N_{\rm H}({\rm cm}^{-2}) = 22$. However,
four high-redshift ($2>z>0.5$) luminous 
($L_{\rm 2-10~keV}=10^{44.5\sim45.5}$ erg s$^{-1}$) broad-line AGNs
(AX~J131816+3240(183), AX~J131054+3004(037), AX~J131724+3203(152), and AX~J131021+3019(039)) are fitted
with large column densities. 
Such large neutral hydrogen column densities conflict with
the existences of broad MgII lines and strong UV continua
in their optical spectra.
However, there is a possibility that optical extinction is not
always strongly correlated with X-ray absorption (e.g., NGC4151), as in the case
where the X-ray absorbing gas is located within the dust sublimation radius,
the absorbing column density is variable, or if the gas-to-dust ratio or the composition 
is different from the Galactic interstellar gas.
These hardness may also be explained by the effect
of the reflection component. The presence of the reflection
component makes observed
type 1~AGN spectrum harder than a photon index of 1.7 in the 0.7--10~keV
band at high-redshift, as shown in Figure \ref{z_pi}.
The slightly hardened X-ray spectrum in
the observed band is spuriously fitted with
large column density in the rest-frame band (1.8--25~keV at a redshift of
1.5) at a high redshift.
To distinguish the effect of a real X-ray absorption and the
reflection component,
X-ray spectroscopic observations for such broad-line AGNs
below 0.7~keV in the observed frame are critical.

\placefigure{lum_nh}

To examine the correlation between the strength of optical broad-line and
the X-ray absorption column density,
we plotted the AGNs with redshift smaller than 0.75 in the
H$ \beta $-to-[OIII] 5007{\AA} versus
hydrogen column density diagram in Figure \ref{RB}.
The narrow- or weak-broad-line AGNs with $\log N_{\rm H}({\rm cm}^{-2}) = 22 \sim 23$
have 10 times smaller values of
H$\beta$/[OIII]5007{\AA} than typical QSOs.
The small ratio suggests existence of
absorption with $A_V$ of larger than 2$\sim$3 mag to the broad-line region.
The optical extinction corresponds to a
hydrogen column density of
$\log N_{\rm H}({\rm cm}^{-2}) = 21.5 \sim 21.7$,
according to the relation
$N_{\rm H}/A_V = 1.79 \times 10^{21}$ cm$^{-2}$~mag$^{-1}$,
which is determined from observations of Galactic objects
(e.g., \cite{pre95}).
Thus, the optical extinction derived from the H$\beta$ region
is consistent with the X-ray spectra for these objects.
One narrow-AGN and 5 weak-broad-line AGNs
with $\log({\rm H}\beta/{\rm [OIII] 5007{\AA}})$ of less than 0 are
fitted with smaller hydrogen column densities 
than $\log N_{\rm H}({\rm cm}^{-2}) = 22$.
In the H$\alpha$ region, however,
3/6 of them have a significant broad H$\alpha$ line
and are plotted with open rectangles or an open star in Figure \ref{RB}.
They fall in the lower-right region in Figure \ref{RAB}.
The small hydrogen column density is consistent with the
existence of a broad H$\alpha$ line, which suggests smaller absorption
to the broad-line region than AGNs with $\log N_{\rm H}({\rm cm}^{-2}) = 22 \sim 23$.
All strong-broad-line AGNs with $\log({\rm H}\beta/{\rm [OIII] 5007{\AA}})$
larger than 0 are fitted with hydrogen column density of less than
$\log N_{\rm H}({\rm cm}^{-2}) = 22$.
These facts suggest the critical column density which
divides the narrow-line and broad-line AGN is 
around $\log N_{\rm H}({\rm cm}^{-2}) = 22$.
The critical column density is consistent with the lower
boundary of the distribution of absorbing column densities of Seyfert 2 galaxies 
(Risaliti, Maiolino, \& Salvati 1999).
Therefore, for AGNs with redshift smaller than 0.75,
there is no broad-line AGN with a large absorption column density and
the strengths of the broad Balmer lines are consistent with
the hydrogen column density derived from the X-ray spectra.

\placefigure{RB}

\section{Number Counts and Contribution to the CXB}
\noindent
We hereafter divide the AGN sample at the intrinsic
column density of $\log N_{\rm H}({\rm cm}^{-2}) = 22$ :
we refer the AGNs with the column density larger and smaller than
$\log N_{\rm H}({\rm cm}^{-2}) = 22$ as ``the absorbed AGN'' and
``the less-absorbed AGN'', respectively.
At first,
we include the four high-redshift broad-line AGNs
fitted with the large column densities 
in the less-absorbed sample,
because such large column densities 
could be only apparent (Section 4.2).

Figure \ref{logN} shows the cumulated logN-logS relations for the absorbed
and for the less-absorbed samples in the SIS $3.5\sigma$ sample.
Since the sensitivity limit is given in count rates, the
actual flux limit depends on X-ray spectrum of sources.
For an X-ray spectrum with the canonical photon index of type~1 AGNs ($\Gamma=1.7$),
the deepest source-detection limit of the SIS $3.5\sigma$ sample
(1.2 cts ksec$^{-1}$; see Table \ref{area}) is equivalent for $1.1 \times 10^{-13}$\fluxunit
in the 2--10~keV band.
If we assume an object with hydrogen column density of 
$\log N_{\rm H}({\rm cm^{-2}})=23$ at redshift of 0.1,
which corresponds to the hardest object in the LSS sample,
the flux limit is estimated to be
$1.6 \times 10^{-13}$\fluxunit in the 2--10~keV band.
In the calculation of Figure \ref{logN}, the SIS
count rate was converted into the flux with the best-fit photon index
determined in the 0.7--10~keV band for each source to take into
account this effect. 
As noted from Figure \ref{logN}, the surface number density
of the absorbed AGNs is comparable to that of the
less absorbed AGNs at flux larger than $2\times10^{-13}$\fluxunit.
The large fraction of the less-absorbed AGNs
in the identified sample is due to
different limiting flux for different X-ray spectra.

\placefigure{logN}

At the first step, we estimate the contributions to the CXB,
integrating the logN-logS relations from
the brightest to the faintest object in each population of the LSS sample.
For less-absorbed AGNs, absorbed AGNs, and clusters of galaxies,
the summed fluxes of sources per unit area
are $3.3\times10^{-12}$\fluxunit degree$^{-2}$,
$6.3\times10^{-13}$\fluxunit degree$^{-2}$,
and $1.4\times10^{-13}$\fluxunit degree$^{-2}$
and contributions to the CXB
are 17\%, 3\%, and 0.7\% , respectively,
if we use the CXB flux determined in the LSS field
(\cite{ish99}, $2.0 \times 10^{-11}$\fluxunit degree$^{-2}$ in the 2--10~keV band).

As the second step, to determine contributions of each population in a wider flux range,
we combined results of the {\it HEAO1\/}~A2 sample (\cite{pic82}).
The sample consists of 30 AGNs,
4 BL Lac objects, 30 clusters of galaxies, 1 starburst galaxy,
and 3 unidentified sources
with the flux limit of $ 3.1 \times 10^{-11} $ \fluxunit\
in the 2--10~keV band.
In the 30 AGNs, 6 AGNs are fitted with hydrogen column density larger than $\log N_{\rm H}({\rm cm}^{-2}) = 22$
(\cite{sch97}).
The source number density at the flux limit is $ 2.4 \times 10^{-3} \
{\rm degree}^{-2} $.
If we assume that there is no break in the logN-logS relations,
the slopes of the relations between
the {\it HEAO1\/}~A2 and the LSS limits can be determined.
The results are shown in Table \ref{tabE}.
In this flux range,
the total number count in the 2--10~keV band has a slope
($\alpha$ of $N(>S) = k S^{-\alpha}$) of
$1.5$ and is consistent with that of smoothly distributed sources in
a simple Euclidean geometry as reported in Ueda et al. (1998).
However, the slope is a summation of
steep slopes for AGNs and
a shallow slope for clusters of galaxies.
Thus the Euclidean distribution is an apparent effect.
The slope of the logN-logS relation of clusters of galaxies
in the flux range is consistent with that 
determined in the 0.5--2~keV band ($1.15$, \cite{vik98}).
Based on the slope of the logN-logS relation,
the contributions from less-absorbed AGNs, absorbed AGNs,
and clusters of galaxies are estimated to be 9\%, 4\%, and 1\%, respectively,
integrating the logN-logS relations from $3 \times 10^{-11}$\fluxunit
down to the flux limit
of $2 \times 10^{-13}$\fluxunit, which is the limit for absorbed AGNs
and at which
16\% of the CXB was resolved into discrete sources.
The logN-logS relation of the absorbed AGNs with the four
high-redshift broad-line AGNs is also shown in Figure \ref{logN} with 
a thin long-dashed line. Based on the logN-logS relation, 
the contribution of the absorbed AGNs with the four
AGNs to the CXB is estimated to be 6\%.

\placetable{tabE}

\section{Redshift Distribution of Identified AGNs and Deficiency of
Obscured QSO}
\noindent
To examine the redshift and luminosity distributions of the AGNs,
we plot the X-ray luminosity versus redshift diagram of 
all the identified AGNs of the SIS $3.5\sigma$ sample
in Figure \ref{sis_lum}a.
Absorbed AGNs are marked with dots. 
At first, the high-redshift broad-line AGNs
which are fitted with large column densities are included in the
less-absorbed sample as in Section 5.
The number ratio of absorbed AGNs-to-less-absorbed AGNs
is clearly changing with redshift, luminosity, or both.
If we limit the sample above the
flux level of $2\times10^{-13}$\fluxunit, which is the
flux limit for the absorbed AGNs,
the redshift and luminosity
distribution of the limited sample is also significantly different
between these two populations.
The number count of the absorbed AGNs
is dominated by nearby
low-luminous objects, in contrast to
that of less-absorbed AGNs, which
are dominated by QSOs at intermediate to high redshifts.
In the same figure, the AGN sample from the {\it HEAO1\/}~A2
survey (\cite{pic82}) is also plotted
with small marks.
Objects with hydrogen column density larger than $\log N_{\rm H}({\rm cm}^{-2}) = 22$ are
also marked with small dots (\cite{sch97}).
The same tendency of deficiency of absorbed AGNs with a large luminosity
is seen in their sample.
Such a deficiency of luminous absorbed AGNs
is also found in the {\it ROSAT}, {\it ASCA}, and
{\it Beppo-SAX} deep surveys in the Lockman Hole (\cite{has99}).
In Figure \ref{sis_lum}b, detection limits of the SIS 2--7~keV 3.5$\sigma$ sample
for intrinsic luminosities with various absorption column densities is shown
as a function of redshift.
In this calculation, we assume the
intrinsic photon index of 1.7 and consider the response function of
{\it ASCA} with the SIS.
The difference between the survey limits
of non-absorbed objects and that of objects with $\log N_{\rm H}({\rm cm}^{-2}) = 22.5$
is less than 0.2 in the logarithmic-scale intrinsic luminosity
at the redshift range from 0 to 1.5. Thus,
we can detect objects
with intrinsic column densities of up to $\log N_{\rm H}({\rm cm}^{-2}) = 22.5$ at redshifts
smaller than 1.5 in an unbiased fashion.
At higher redshifts
objects with column densities of up to $\log N_{\rm H}({\rm cm}^{-2}) = 23$
are detected without bias, thanks to the redshift effect.
It should be noted that the peak of the distribution of absorption column densities
of type~2 Seyferts in the nearby universe ($\log N_{\rm H}({\rm cm}^{-2}) > 23$; \cite{ris99})
is beyond the detection limit.

\placefigure{sis_lum}

To examine the deficiency of luminous absorbed AGNs more quantitatively,
we compare the redshift distribution of absorbed AGNs and
that of less-absorbed AGNs, considering the difference of their X-ray spectra.
In Figure \ref{zdis}a, we show the redshift distribution of
less-absorbed AGNs
as well as the expected redshift distribution of broad-line AGNs.
The four high-redshift broad-line AGNs
which are fitted with large column densities are included in the
less-absorbed sample.
The expected redshift distribution of broad-line AGNs
is calculated based on a hard-band X-ray luminosity function of broad-line AGNs
and its evolution (\cite{boy98a});
they are determined by an AGN sample from
{\it ASCA} observations of Deep {\it ROSAT} fields and optical identifications of the Large Area Sky Survey/Modulation
Collimator catalog obtained from {\it HEAO1} mission.
The expected redshift distribution matches well to
that of the LSS sample.
Based on Kolmogorov-Smirnov test, the 
probability that the observed redshift distribution agrees with the model
is calculated to be 64\%. 
Thus, we use the hard-band X-ray luminosity function as a base model
for less-absorbed AGNs.
In Figure \ref{zdis}b, we compare the redshift distribution of
absorbed AGNs in the LSS sample
with the expected distribution of them
calculated by
assuming that the shape of the intrinsic (which means absorption corrected)
luminosity function of absorbed AGNs is the
same as that of less-absorbed AGNs.
The equal number density of AGNs with $\log N_{\rm H}({\rm cm}^{-2})=22\sim23$ to that
of less-absorbed AGN
is the same as in the spectrum model of the CXB;
e.g., Comastri et al. (1995) used the value of 1.23 for the number ratio of
AGNs with absorption of $\log N_{\rm H}
({\rm cm}^{-2})= 22\sim23$ to those with absorption less than $\log N_{\rm H}({\rm cm}^{-2})=22$.
As seen in Figure \ref{zdis}b, 
at redshift smaller than 0.4, in which the X-ray luminosity of the sample
AGN is less than $10^{44}$ erg s$^{-1}$,
the number of absorbed AGNs is comparable to the expectation.
However, at redshifts larger than 0.4, the total expected number is about 10,
which contrasts to the non-detection.
The observed redshift distribution is different from 
the model redshift distribution; Kolmogorov-Smirnov test gives
a probability of only  5\% for the null hypothesis that the
redshift distribution of absorbed AGNs is the same as 
the model, thus the hypothesis is rejected.  
Therefore, our sample suggests the deficiency of the absorbed AGNs
with the column density of $\log N_{\rm H}({\rm cm}^{-2})=22\sim23$,
in the redshift range between 0.5 and 2.0, or
in the X-ray luminosity range larger than $10^{44}$ erg s$^{-1}$, or both.
We plot the redshift distribution of the four high-redshift
broad-line AGNs with hard X-ray spectra in Figure \ref{zdis}b
with a histogram shaded with dashed lines.
By including the four AGNs, the observed redshift distribution agrees 
the model redshift distribution with probability of 60\%.
If their large column densities are real,
they could complement the deficiency of absorbed high-redshift luminous AGNs.
Revealing the column density distribution of absorbed AGN as a function of
redshift and luminosity and constructing the model of the origin of the
CXB more accurately will be an important objective for the next generation X-ray surveys.

\placefigure{zdis}

\section{Properties of identified AGNs in other wavelengths}
\subsection{Optical Photometric Properties}
\noindent
We have examined the optical appearance of each identified object.
The results are indicated in the last column of the Table \ref{flux_table}.
All of the low-luminosity AGNs show their host galaxy component in the optical light.
We calculate optical total absolute magnitudes of the identified objects,
using $\alpha=-0.5$, which corresponds to $V-R$ of 0.22 mag.
The results are shown in Table \ref{flux_table} and
plotted in Figure \ref{opt_x3} as a function of X-ray luminosity.
For high-luminosity AGNs, hard X-ray luminosities correlate well with optical luminosities.
On the other hand, for the low-luminosity AGNs whose optical light is
affected and maybe dominated by their host galaxy components, the correlation is broken. Considering the break, we estimate 
the absolute magnitudes of the host galaxies to be 
around $M_V=-21\sim-23.5$ mag and
are similar to that of QSO host galaxies (\cite{bah97}).

\placefigure{opt_x3}

The optical colors of luminous broad-line AGNs distribute from $B-R$ of $-0.2$
mag to $1.1$ mag,
consistent with the expected color range for broad-line AGNs.
There is no broad-line QSO which has a $B-R$ color larger than $2.0$
mag, which is a criterion for {\it red} QSOs in Kim \& Elvis (1999).
Two low-luminosity broad-line AGNs 
(AX~J131725+3300(192) and AX~J131407+3158(127))
which show a broad H$\alpha$ emission
have red optical colors. 
The red colors are probably affected by the host galaxy,
because their optical images 
show the host galaxy components.

\subsection{Radio and Far-infrared Properties}
\noindent
To examine radio properties of the identified AGNs,
we show a radio versus hard X-ray luminosity diagram in Figure \ref{radio_x1}.
There are two sequences in the diagram;
three objects which have large radio-to-hard X-ray luminosity ratios
correspond to radio-loud AGNs and others which have smaller
ratios correspond to radio-quiet AGNs.
The radio-to-hard X-ray ratios of
the radio-loud sequence are consistent
with the average SED of radio-loud QSO (\cite{elv94}).
On the other hand,
the radio-quiet sequence has one order of magnitude larger
radio-to-hard X-ray luminosity ratio than the average
SED of radio-quiet QSO (\cite{elv94}).
Based on the dispersion of the radio-to-optical luminosity
ratios of radio-quiet QSOs (\cite{vis92}),
the large radio-to-hard X-ray luminosity ratio is
accounted by a scatter and a selection bias for radio louder objects.
The fraction (3/30) of radio-loud AGNs in the total AGN sample is consistent with
that in the low-luminosity sample of optically selected QSOs (13\%)
(\cite{vis92}).
In radio images, all radio-quiet AGNs show only one point-like
component. Two radio-loud AGNs consist of two components;
one dominates the total flux located on the optical position and the
other fainter source
is $\sim15$ arcsec away from the central component.

\placefigure{radio_x1}

In the LSS sample, only the galactic star is detected in the {\it IRAS} survey (\cite{mos92}).
None of the AGN is detected in the {\it IRAS} Faint Source Survey.
Based on the flux limit of the {\it IRAS} survey in the LSS region
(0.1 Jy in 60 $\mu$m),
most of the AGNs have lower limits to the logarithmic ratio
of 2--10~keV luminosity-to-far-infrared luminosity ($\nu_{60\mu{\rm m}}
L\nu_{60\mu{\rm m}}$)
of less than $-1$,
consistent with those known for QSOs ($-1$) and
only one object (AX~J131822+3347 (235)) has a lower limit ($-0.6$)
to the logarithmic ratio larger than the typical value of QSOs.
The lower limits to the logarithmic ratio are significantly
larger than those of star-forming galaxies,
which have logarithmic ratios around $-4$.

\section{Summary}
\noindent
We present results of optical identifications
of the X-ray sources detected in the {\it ASCA} Large Sky Survey.
Optical spectroscopic observations were done for
34 X-ray sources which were detected with the SIS in the 2--7~keV band
above 3.5 $\sigma$. The flux limit corresponds to
$\sim 1\times10^{-13}$ \fluxunit in the 2--10~keV band.
The sources are identified with 30 AGNs,
2 clusters of galaxies, and 1 galactic star.
Only 1 source is still unidentified.

All of the X-ray sources that have a hard X-ray spectrum
with an apparent photon index of smaller than 1 in the 0.7--10~keV
band
are identified with narrow-line or weak-broad-line AGNs
at redshifts smaller than 0.5.
This fact supports the idea that absorbed X-ray spectra of narrow-line and
weak-broad-line AGNs make
the Cosmic X-ray Background (CXB) spectrum
harder in the hard X-ray band than that of a broad-line AGN,
which is the main contributor in the soft X-ray band.
Assuming their intrinsic spectra are 
same as a broad-line AGN (a power-law model with a photon index of 1.7),
their X-ray spectra are fitted with
hydrogen column densities of $\log N_{\rm H}({\rm cm}^{-2}) = 22 \sim 23$
at the object's redshift.
On the other hand,
X-ray spectra of the other AGNs are consistent with that of a
nearby type~1 Seyfert.
In the sample, four high-redshift luminous broad-line AGNs show a hard X-ray spectrum
with an apparent photon index of $1.3\pm0.3$.
The hardness
may be explained by the reflection component of a type~1 Seyfert.
The hard X-ray spectra may also be explained by absorption with
$\log N_{\rm H}({\rm cm}^{-2}) = 22 \sim 23$ at the object's redshift,
if we assume an intrinsic photon index of 1.7.
The origin of the hardness is not clear yet.

Based on the logN-logS relations of each population,
contributions to the CXB in the 2--10~keV band are estimated to be
9\% for less-absorbed AGNs ($\log N_{\rm H}({\rm cm}^{-2}) < 22$)
including the four high-redshift broad-line AGNs with a hard X-ray spectrum,
4\% for absorbed AGNs ($22 < \log N_{\rm H}({\rm cm}^{-2}) < 23$, without
the four hard broad-line AGNs), and
1\% for clusters of galaxies in the flux range from
$3\times10^{-11}$\fluxunit to $2\times10^{-13}$\fluxunit.
If the four hard broad-line AGNs are included in the absorbed AGNs,
the contribution of the absorbed AGNs to the CXB is estimated to be 6\%.

In optical spectra, there is no high-redshift
luminous cousin of a narrow-line
AGN in our sample. The redshift distribution of the absorbed AGNs
are limited below $z=0.5$ excluding the four hard broad-line AGNs,
in contrast to the existence of 15
less-absorbed AGNs above $z=0.5$.
The redshift distribution of the absorbed AGNs
suggests a deficiency of AGNs with column densities of
$\log N_{\rm H}({\rm cm}^{-2}) = 22$ to $23$ in the redshift range between 0.5
and 2, or in the X-ray luminosity range
larger than $10^{44}$ erg s$^{-1}$,
or both.
If the large column densities of the four hard broad-line AGNs are real, 
they could complement the deficiency of 
X-ray absorbed luminous high-redshift AGNs.

\acknowledgments{
MA, KO, and TY would like to thank S.Okamura, M.Sekiguchi and
the MOSAIC CCD camera team, and staff members of the KISO observatory for
their support during the imaging observations.
MA, KO, and TY appreciate the support from members of the 
University of Hawaii observatory
during the spectroscopic observations.
KO is grateful to the hospitality during his stay at the 
Institute for Astronomy, University of Hawaii,
where a part of this work was done.
MA, KO, YU, and IL wish to thank members of the Calar Alto observatory for
their help during the spectroscopic observations.
We are also grateful to the referee for his useful suggestions.
This research has made use of NASA/IPAC Extragalactic Database (NED),
which is operated by the Jet Propulsion Laboratory, Caltech,
under contract with the National Aeronautics and Space Administration.
MA, WK, and MS acknowledge support from a
Research Fellowships of the Japan Society for the Promotion of Science
for Young Scientists.
The optical follow-up program is supported by grants-in-aid from the
Ministry of Education, Science, Sports and Culture of Japan
(06640351, 08740171, 09740173) and from the Sumitomo Foundation.
}

\appendix
\section{{\it ROSAT} HRI Data Reduction and Source Detection}
\noindent
{\it ROSAT} HRI observations were made in 2 fields with a 20 ksec exposure
in December 1997.
These fields were centered at $\alpha$=13$^{\rm h}$14$^{\rm m}36^{\rm s}$,
$\delta$=32$^{\circ}$01$'$12$''$ (LSS-HRI1) and $\alpha$=13$^{\rm h}$12$^{\rm m}29^{\rm s}$,
$\delta$=31$^{\circ}$13$'$12$''$ (LSS-HRI2) (J2000) with a radius of $19'$.

These data were reduced with EXSAS package (\cite{zim94})
on the MIDAS.
For the source detection, we applied the
{\bf detect} command selecting photons 
in the PHA channel range from 2 to 8, only.
X-ray sources detected in the HRI follow-up observations
with Maximum Likelihood larger than 10 were listed in Table \ref{hri_list}.

The conversion factor from a count rate to a 0.5--2.0~keV flux is
$1.854 \times 10^{-14}$ \fluxunit cts$^{-1}$ ksec.
Thus the flux of the faintest source corresponds to $1.13 \times 10^{-14}$ \fluxunit.
Considering systematic shifts between candidates of optical
counterparts and X-ray sources, we shifted the X-ray positions
with $\Delta \alpha = -3.\!^{\prime\prime}39$, $\Delta \delta = +5.\!^{\prime\prime}08$ and
$\Delta \alpha = -2.\!^{\prime\prime}37$, $\Delta \delta = +3.\!^{\prime\prime}18$ for regions LSS-HRI1 and
LSS-HRI2,
respectively.

\placetable{hri_list}

\section{Cross identification with the {\it ROSAT} PSPC catalogs and
the VLA FIRST catalog}
\noindent
We summarize the identifications of
the {\it ASCA} LSS sources with the {\it ROSAT} PSPC
sources and the VLA FIRST radio sources in Table \ref{corre_table}.

\placetable{corre_table}

\clearpage

\figcaption[f1a.eps,f1b.eps,f1c.eps,f1d.eps,f1e.eps,f1f.eps,
f1g.eps,f1h.eps,f1i.eps]{Optical finding charts and spectra of
identified objects.
The field of view of the finding charts is
$1.\!^{\prime}6 \times 1.\!^{\prime}6$. Large circles have a radius
of $0.\!^{\prime}6$ and correspond to the 90\% confidence
error circles of ASCA sources.
Middle and small size circles have radii of $20''$ and $5''$
and correspond to error circles of {\it ROSAT} PSPC and HRI sources,
respectively.
Positions of FIRST radio sources are indicated by small rectangles.
Objects which have an UV-excess or a blue $B-R$ colors are marked
with pentagons. 
The objects listed in Table 2 are labeled with ``A'', ``B'', 
``C'', and ``Z''.
The spectrum of each identified object is shown
with the identification label and classification.
Positions of atmospheric absorption lines are shown as a
cross with a circle. \label{plate}}

\figcaption[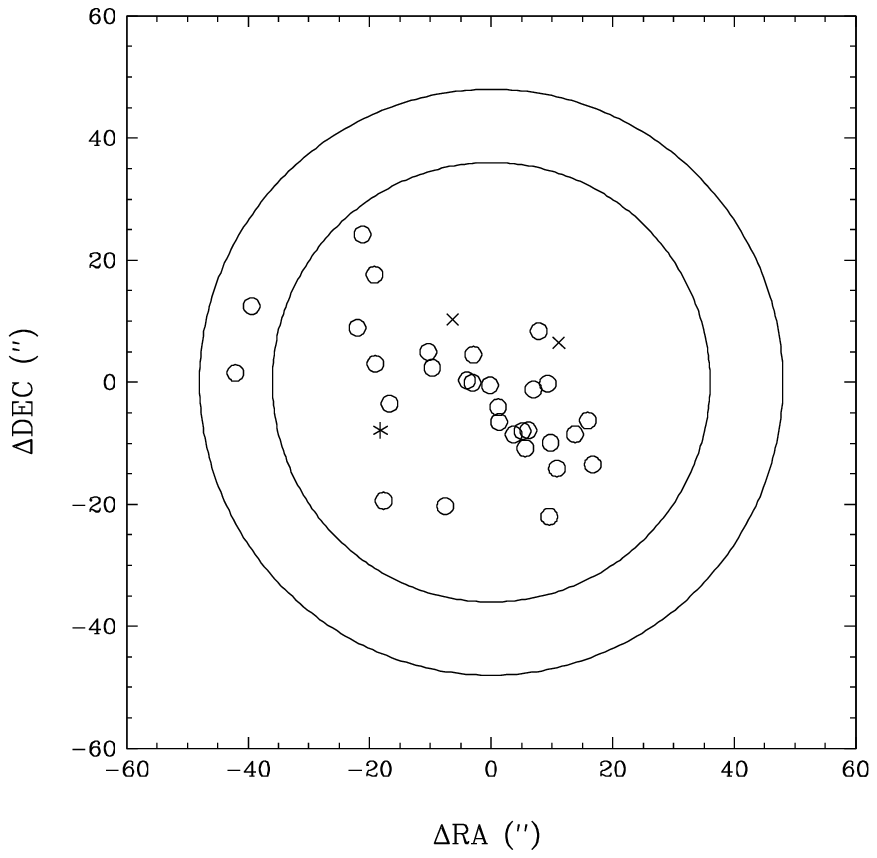]{The distribution of the identified objects within
the X-ray error circles.
Open circles, crosses, and an asterisk
represent AGNs, clusters
of galaxies, and a galactic star, respectively.
Inner and outer circles have radii of $0.\!^{\prime}6$
and $0.\!^{\prime}8$.
The inner circle corresponds to the estimated error radius of the
LSS X-ray sources in the 90\% confidence level. \label{coord}}

\figcaption[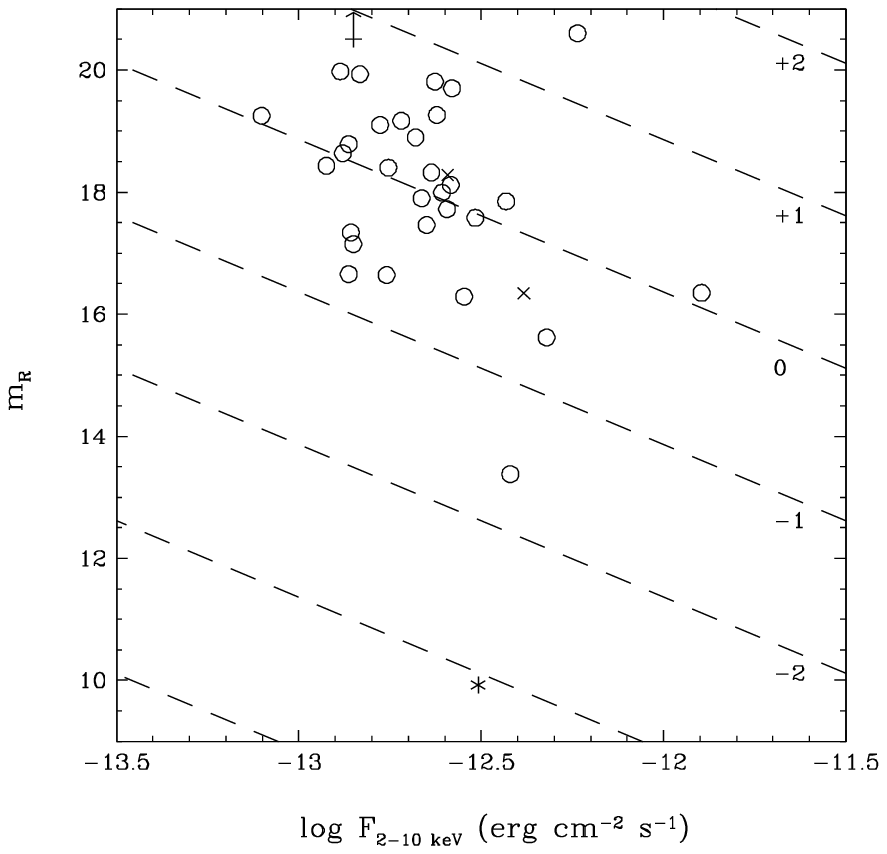]{2--10~keV flux versus optical $R$ band magnitude 
diagram of the identified objects.
Marks are the same as in Figure 2. The upper limit of the optical magnitude of AX~J131832+3259(199)
is indicated with a plus sign with an upward arrow.
Dashed lines represent X-ray-to-optical flux ratio of
$\log f_{\rm X}/f_{\rm V} = +2,+1,0,-1,-2,-3,$ and $-4 $
from top to bottom. \label{opt_x1}}

\figcaption[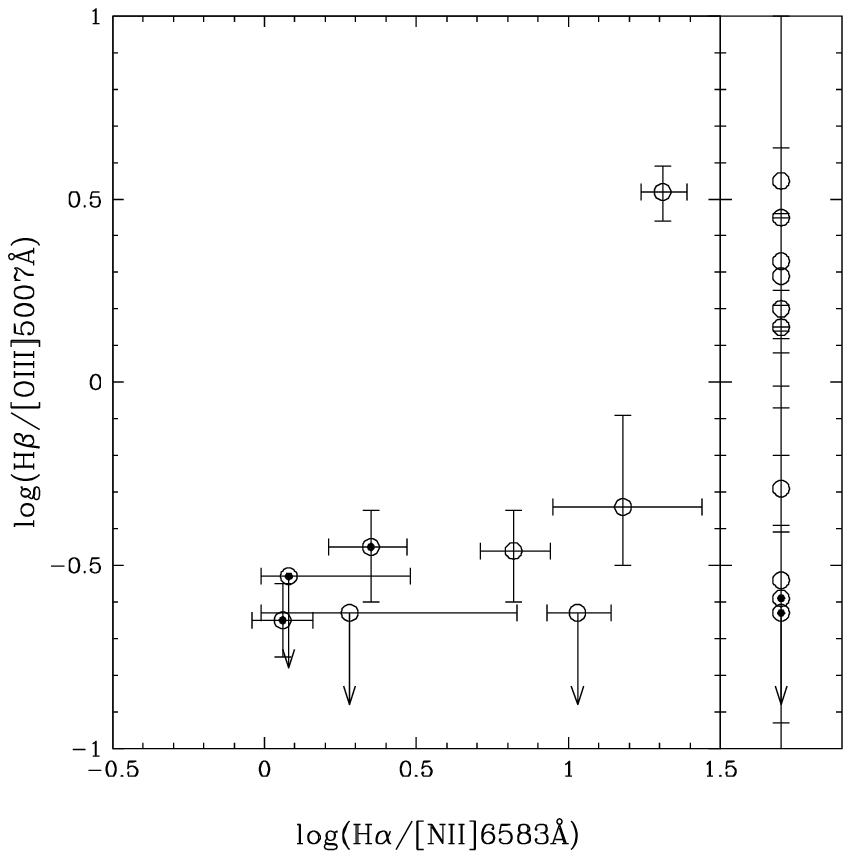]{The distribution of identified AGNs in the
$\log $(H$\alpha $/[NII]6583{\AA}) 
versus $\log $(H$\beta $/[OIII]5007{\AA}) diagram.
Equivalent widths for Balmer lines include both broad and narrow
emission lines.
Objects which were not observed in the H$\alpha $ region are plotted in the
right box. Upper limits are indicated with downward arrows.
AGNs which have a hard X-ray spectrum (an apparent photon index smaller than 1)
are indicated with dots. \label{RAB}}

\figcaption[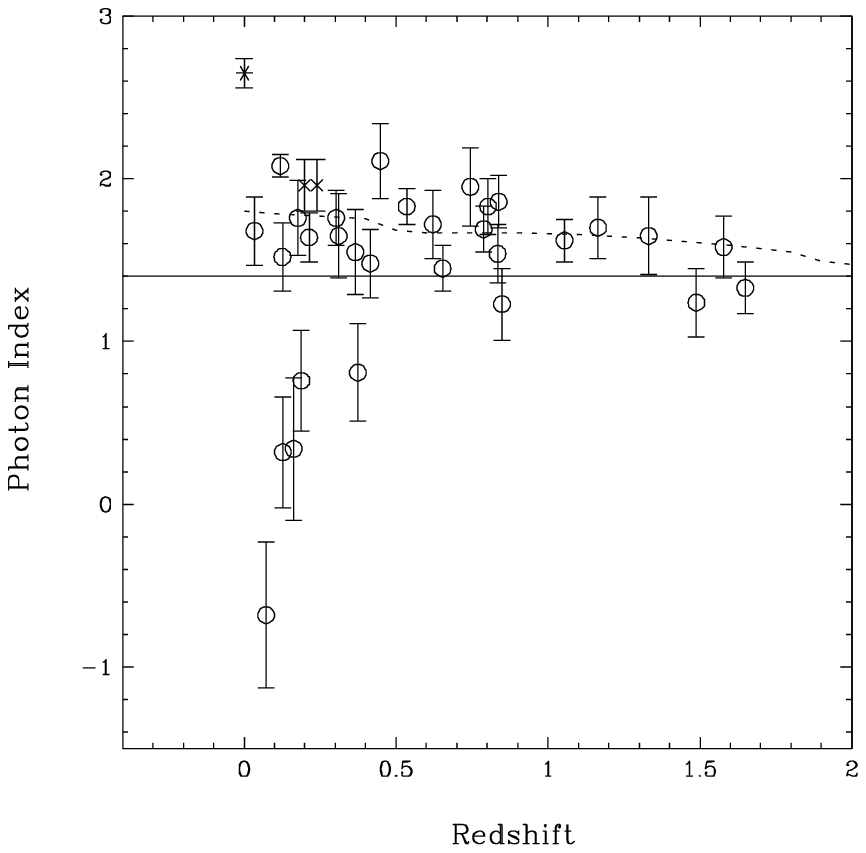]{Apparent photon index (0.7--10~keV) versus
redshift diagram of the identified objects.
Marks are the same as in Figure 2.
Horizontal solid line shows the photon index of the CXB.
Expected apparent photon index for a type~1 Seyfert in the
observed 0.7--10~keV band is shown with a dashed line as
a function of redshifts. See Section 4. \label{z_pi}}

\figcaption[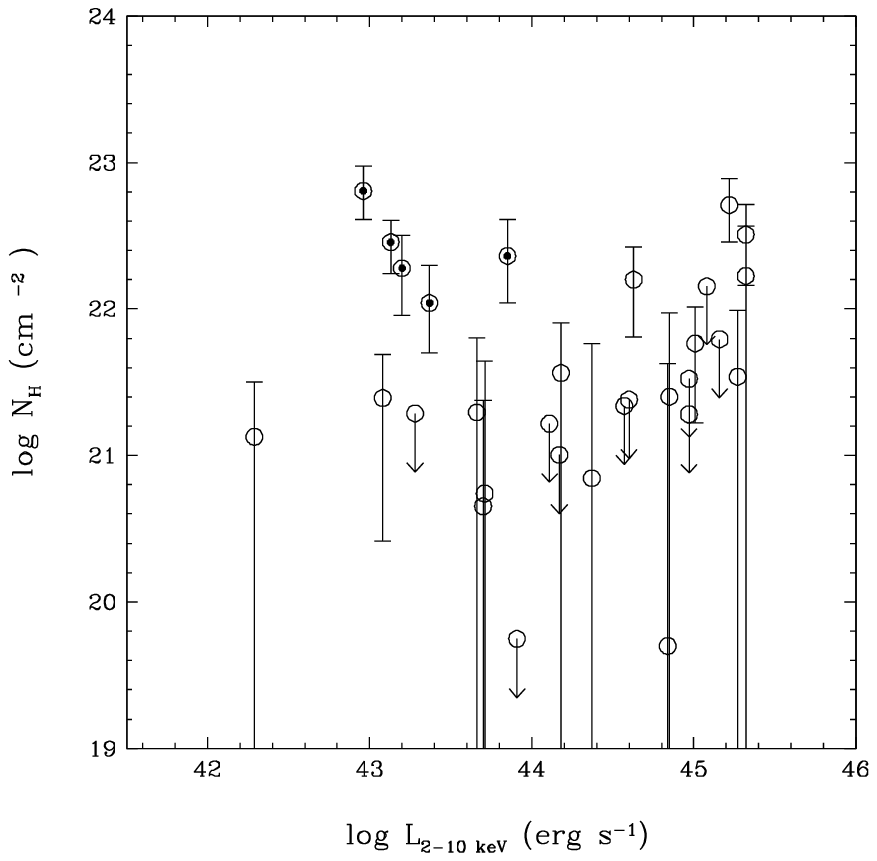]{Fitted column density versus 2--10~keV luminosity diagram
of the identified AGNs.
The X-ray luminosities are not corrected for the absorption.
AGNs with a hard X-ray spectrum which have an apparent
photon index smaller than 1
are indicated with dots.
Upper limits for column densities are indicated with
downward arrows. \label{lum_nh}}

\figcaption[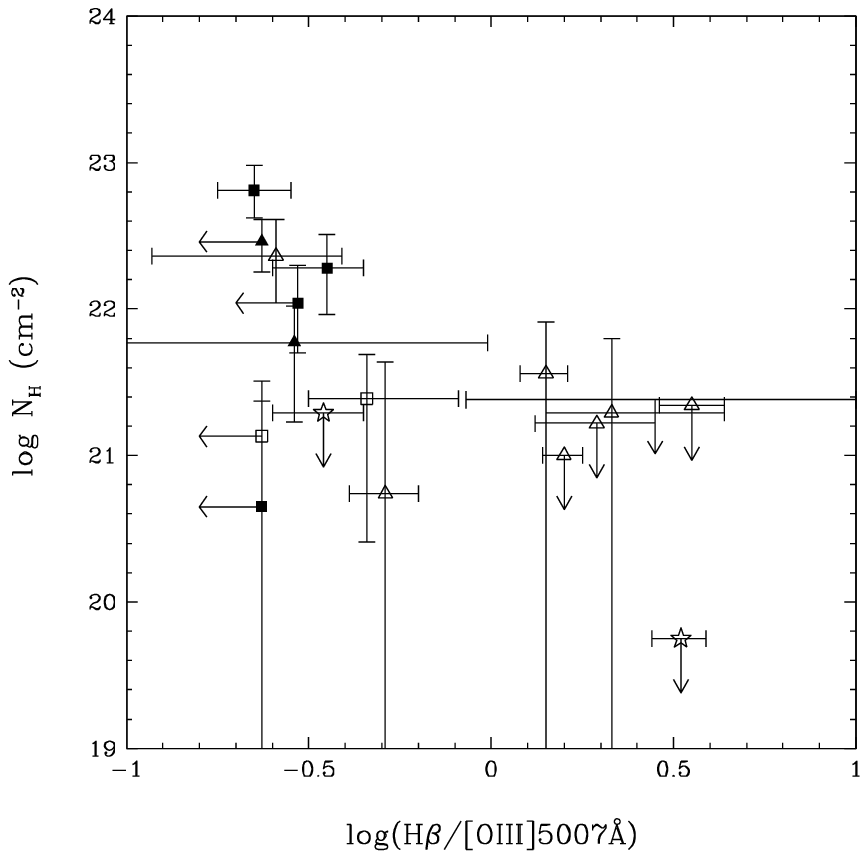]{Fitted column density versus log(H$\beta $ /[OIII] 5007\AA)
of the identified AGNs.
An AGN which shows a broad H$\alpha$ and a broad H$\beta$ lines is
shown with an open star, AGNs with a broad H$\alpha$ and no significant
broad H$\beta$ line are plotted with open rectangles, and AGNs without
significant broad
H$\alpha$ nor H$\beta$ are indicated with filled rectangles.
Note these AGNs are observed in both H$\alpha$ and H$\beta$
regions.
AGNs which are observed only in the H$\beta$ region are plotted
with open triangles for objects with a broad H$\beta$ line,
and with filled triangles for objects without broad H$\beta$ line.
Upper limits are indicated with downward and
leftward arrows. \label{RB}}

\figcaption[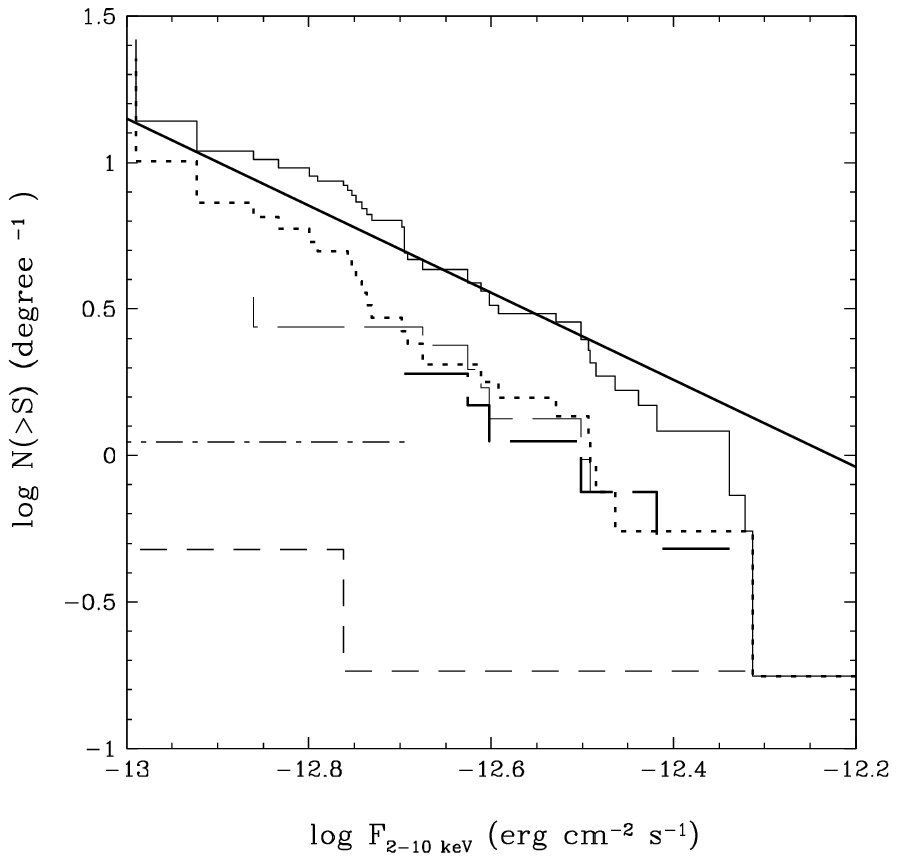]{Cumulative logN-logS relations in the
2--10~keV band for the SIS 2--7~keV 3.5$\sigma$ sample.
The thin solid line, the thick dotted line, the thick long-dashed
line, and the thin long-dashed line correspond to
the logN-logS relations for
all sources, the less-absorbed 
($\log N_{\rm H}({\rm cm}^{-2}) < 22$)
AGNs, the absorbed
($\log N_{\rm H}({\rm cm}^{-2}) > 22$ without the 4 hard broad-line
AGNs) 
AGNs, and the absorbed  
($\log N_{\rm H}({\rm cm}^{-2}) > 22$ with the 4 hard broad-line
AGNs) AGNs, respectively.
The short dashed line indicates the logN-logS relations of clusters
of galaxies, and the dot dashed line shows that of an unidentified source.
The thick solid line represents extrapolation with a slope
of $-1.5$ from the {\it HEAO1\/}~A2 results.
Survey limits for less-absorbed AGNs and clusters of galaxies
are $1 \times 10^{-13}$\fluxunit, whereas it is
$1.6 \times 10^{-13}$\fluxunit for absorbed AGNs with the hardest
X-ray spectrum (see Section 5).\label{logN}}

\figcaption[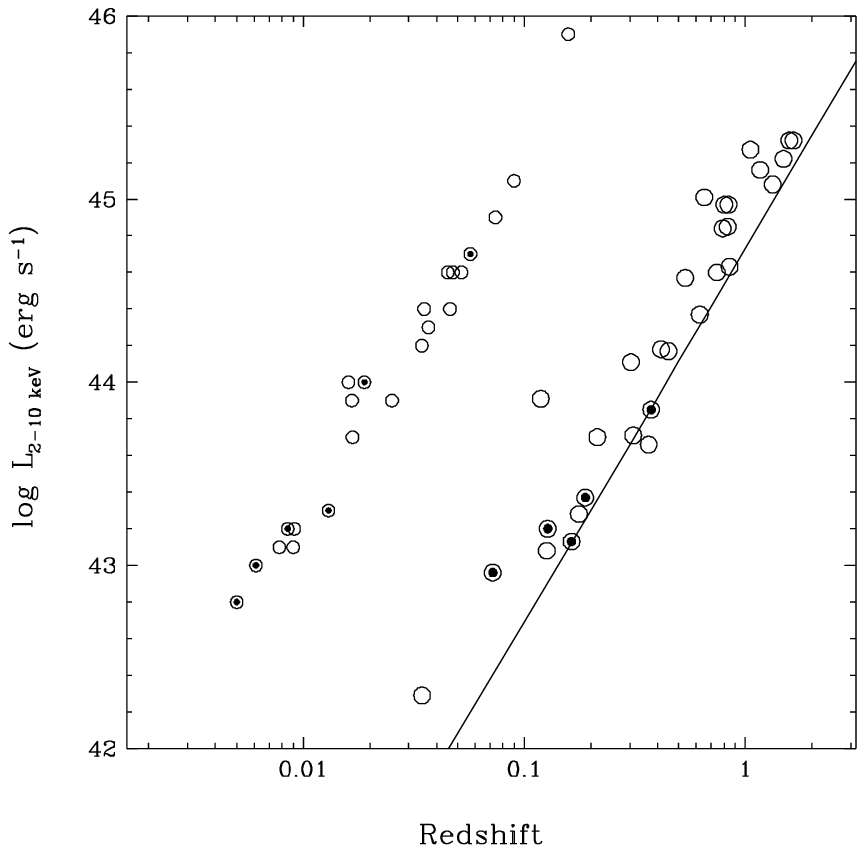, 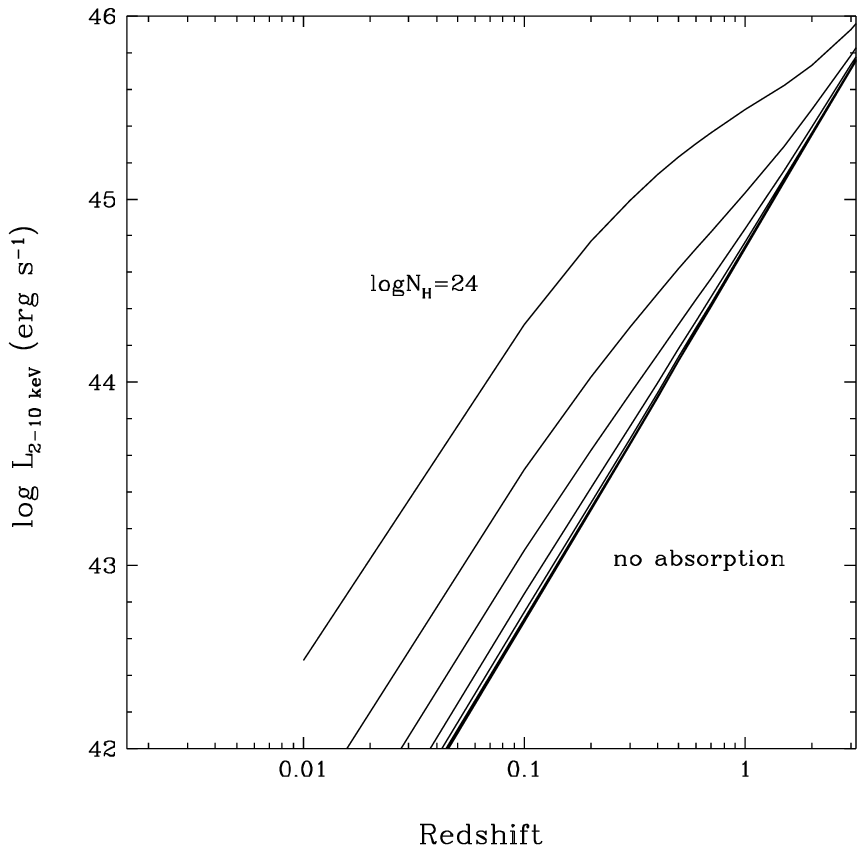]{
a) 2--10~keV luminosity versus redshift diagram of the identified sample.
The LSS AGNs are plotted with large open circles.
Absorbed ($\log N_{\rm H}({\rm cm}^{-2}) > 22$) AGNs are marked with dots.
The solid line is the detection limit of the SIS
2--7~keV 3.5$\sigma$ sample for non-absorbed AGN.
We also plot the {\it HEAO1\/}~A2 sample of AGNs (Piccinotti et al. 1982)
with small marks.
b) Detection limit of the SIS 2--7~keV 3.5$\sigma$ sample to
an {\it intrinsic} 2--10~keV luminosity as a function of redshift.
From top to bottom, column densities
of $\log N_{\rm H}({\rm cm}^{-2})$ =
24.0, 23.5, 23.0, 22.5, 22.0, 21.5, 21.0, and
no absorption with an intrinsic photon index of 1.7 are assumed.
The survey limit of the SIS 3.5$\sigma$ sample is 1.2 cts ksec$^{-1}$
in the 2--7 keV band.
\label{sis_lum}}

\figcaption[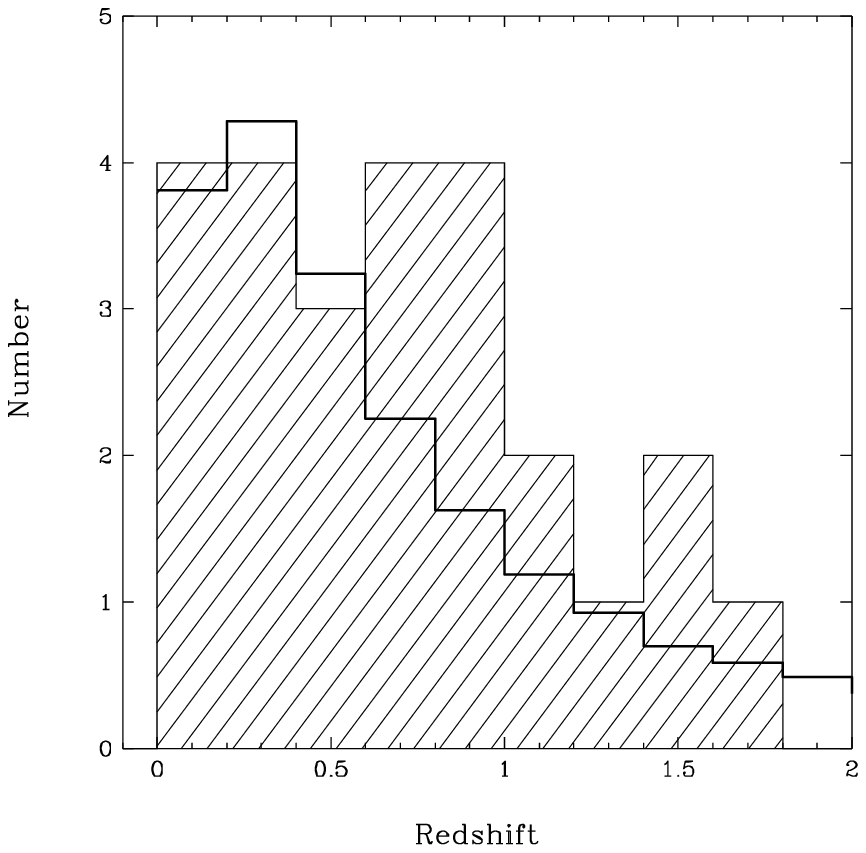, 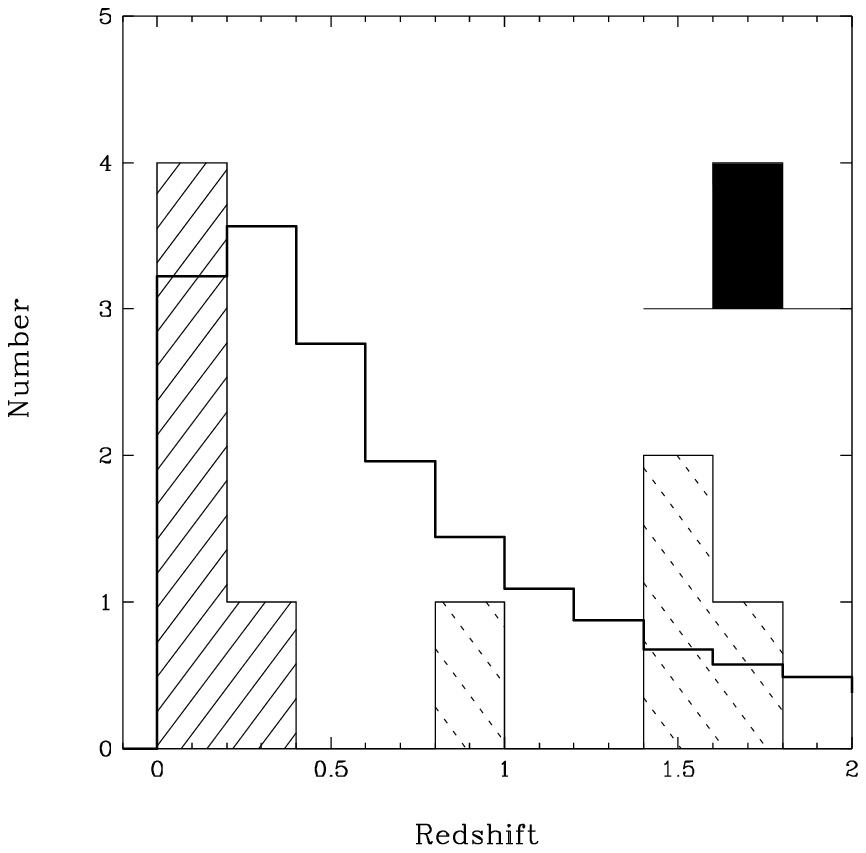]{
a) Comparison of redshift distribution of the
less-absorbed ($\log N_{\rm H}({\rm cm}^{-2}) < 22$)
AGNs in the LSS
sample with that expected from a broad-line AGN luminosity function
in the hard X-ray band.
The shaded histogram is the redshift distribution of the LSS AGN
sample and the solid line is the expected redshift distribution.
The four high redshift broad line AGNs with a hard X-ray spectrum
are included in this histogram. See Section 6.
b) Comparison of redshift distribution of the absorbed 
($\log N_{\rm H}({\rm cm}^{-2}) > 22$)
AGNs in the LSS sample with that expected from
a broad-line AGN luminosity function in the hard X-ray band
with an absorption of $\log N_{\rm H}({\rm cm}^{-2})=22.5$.
The shaded histogram is the redshift distribution of the absorbed
AGNs and the histogram shaded with dashed lines is that of the broad-line
AGNs with a hard X-ray spectrum mentioned above.
The solid line is the expected redshift distribution.
The black histogram at the right-top corner refers to
unidentified source.
\label{zdis}}

\figcaption[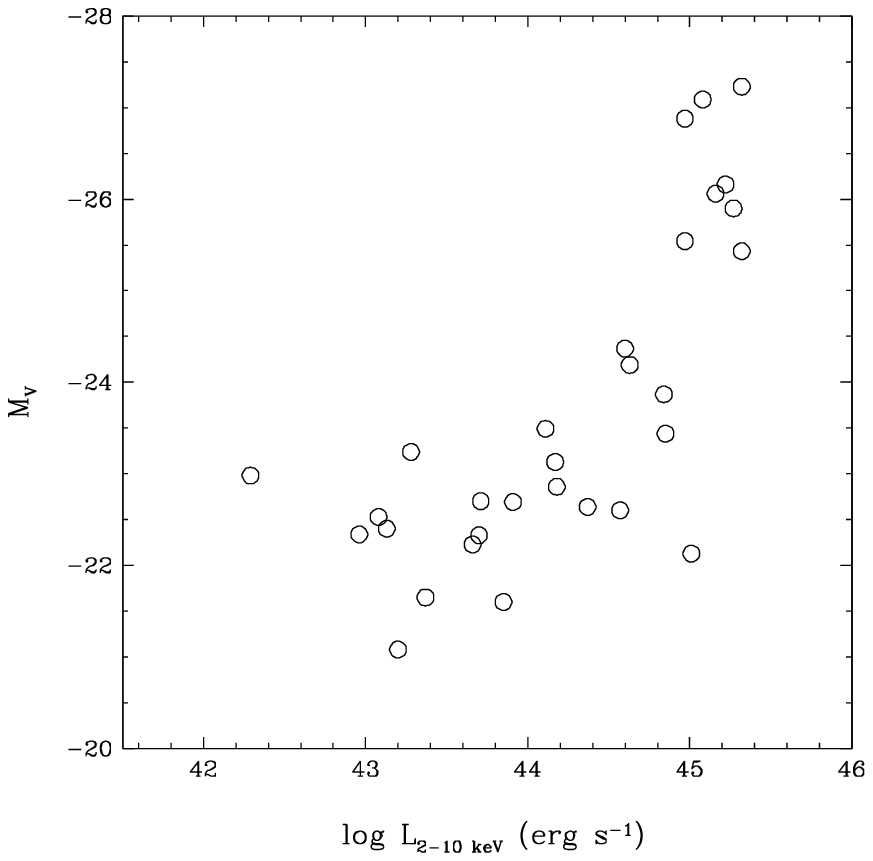]{Hard X-ray luminosity versus optical $V$ band absolute magnitude diagram 
of the identified AGNs.
\label{opt_x3}}

\figcaption[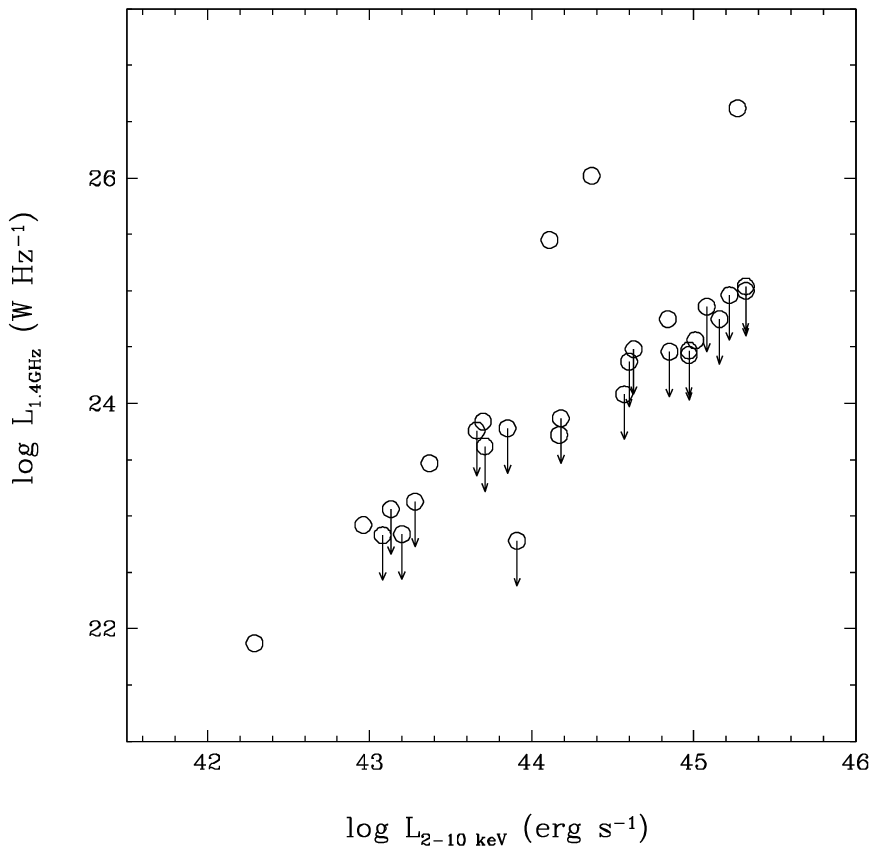]{Hard X-ray luminosity versus radio 1.4 GHz luminosity diagram of the identified AGNs.
Downward arrows show upper limit on the radio luminosities.
\label{radio_x1}}

\clearpage

\begin{table*}
\begin{center}
\begin{tabular}{rcc} \hline \hline
\multicolumn{2}{c}{Count Rate} & \multicolumn{1}{c}{Area} \\
\multicolumn{2}{c}{(cts ksec$^{-1}$)} & \multicolumn{1}{c}{(degree$^2$)} \\
\hline
1.2   &&   0.5    \\
1.4   &&   1.0    \\
1.6   &&   2.0    \\
2.0   &&   3.0    \\
3.0   &&   4.8    \\
4.0   &&   5.0    \\
6.0   &&   5.5    \\
10.0  &&   5.7    \\
20.0  &&   5.8    \\ \hline
\end{tabular}
\end{center}
\tablenum{1}
\caption{Survey Area as a function of the SIS 2--7~keV count rate for the
3.5$\sigma$ sample from Paper I.\label{area}}
\end{table*}

\begin{table*}
\tablenum{2}
\dummytable\label{id_table}
\end{table*}

\begin{table*}
\tablenum{3}
\dummytable\label{spe_table}
\end{table*}

\begin{table*}
\tablenum{4}
\dummytable\label{flux_table}
\end{table*}

\begin{table*}
\begin{center}
\begin{tabular}{lrrrr} \hline \hline
                            &\multicolumn{2}{c}{log N($>$S)\tablenotemark{b}} & \multicolumn{1}{c}{$\alpha$\tablenotemark{c}}      &  \multicolumn{1}{c}{Contributions\tablenotemark{d}}        \\
                            &  {\it HEAO1\/}~A2   & {\it ASCA} LSS   &           &     \\  \hline
Flux limit\tablenotemark{a} & $3\times10^{-11}$ & $2\times10^{-13}$ &       & $2\times10^{-13}$  \\ \hline
Total                      &  $-2.62$         & $  0.80^{+0.07}_{-0.08} $    & $ 1.55^{+0.03}_{-0.05}  $  & 16\%          \\
less-absorbed AGN          &  $-2.66$         & $  0.47^{+0.08}_{-0.10} $    & $ 1.42^{+0.04}_{-0.05}  $  &  9\%           \\
absorbed AGN (without 4 broad-line AGN) &  $-3.68$         & $  0.28^{+0.16}_{-0.25} $    & $ 1.80^{+0.07}_{-0.11}  $  &  4\%          \\
absorbed AGN (with 4 broad-line AGN)    &  $-3.68$         & $  0.44^{+0.13}_{-0.19} $    & $ 1.89^{+0.06}_{-0.09}  $  &  6\%      \\    
Cluster                    &  $-2.98$         & $ -0.74^{+0.23}_{-0.52} $    & $ 1.02^{+0.11}_{-0.24}  $  &  1\%        \\ \hline
\end{tabular}
\end{center}
\tablenotetext{a}{Flux limit in unit of \fluxunit in the 2--10~keV band.}
\tablenotetext{b}{Source number densities at the flux limits in
 degree$^{-2}$. The errors of the LSS source number densities are
estimated with the square root of the detected number.}
\tablenotetext{c}{Slopes of logN-logS relations in the flux range.}
\tablenotetext{d}{Contributions to the CXB from the sources in the flux range from $3\times10^{-11}$
\fluxunit to the flux limits.}
\tablenum{5}
\caption{LogN-logS relations of various populations and their 
contributions to the CXB
in the 2--10~keV band.\label{tabE}}
\end{table*}

\begin{table*}
\tablenum{6}
\dummytable\label{hri_list}
\end{table*}

\begin{table*}
\tablenum{7}
\dummytable\label{corre_table}
\end{table*}

% \clearpage
% \plotone{plate.eps}

\clearpage
\plotone{f2.eps}

\clearpage
\plotone{f3.eps}

\clearpage
\plotone{f4.eps}

\clearpage
\plotone{f5.eps}

\clearpage
\plotone{f6.eps}

\clearpage
\plotone{f7.eps}

\clearpage
\plotone{f8.eps}

\clearpage
\plotone{f9a.eps}

\clearpage
\plotone{f9b.eps}

\clearpage
\plotone{f10a.eps}

\clearpage
\plotone{f10b.eps}

\clearpage
\plotone{f11.eps}

\clearpage
\plotone{f12.eps}

\end{document}